\documentclass[amssymb,aps,twocolumn,floats,showpacs]{revtex4-1}
\usepackage{color,graphicx,pstricks,float}
\usepackage{natbib}

\usepackage{fancybox}
\usepackage{epsfig}
\usepackage{graphicx}
\usepackage{tabularx}
\usepackage{inputenc}
\usepackage{amsmath}
\usepackage{amssymb}
\usepackage{gensymb}
\usepackage{array}

\begin{document}

\title{New Microscopic Magnetic Hamiltonian for Exotic Spin Textures in Metals}

\author{Deepak S. Kathyat}
\author{Arnob Mukherjee}
\author{Sanjeev Kumar}

\address{ Department of Physical Sciences,
Indian Institute of Science Education and Research (IISER) Mohali, Sector 81, S.A.S. Nagar, Manauli PO 140306, India \\
{\it sanjeev@iisermohali.ac.in}
}

\begin{abstract}
We derive a new microscopic spin Hamiltonian for Rashba-coupled double exchange metals.
The Hamiltonian consists of anisotropic interactions of the Dzyaloshinskii-Moriya (DM) and Kitaev form, in addition to the standard isotropic term. 
We validate the spin Hamiltonian by comparing results with those on the exact spin-fermion model, and present its phase diagram using large scale Monte Carlo simulations. In addition to ferromagnetic, planar spiral and flux states, the
model hosts skyrmion crystal and classical spin-liquid states characterized, respectively, by multiple peaks and a diffuse ring pattern in the spin structure factor.
The filamentary domain wall structures in the spin-liquid state are in remarkable agreement with experimental data on thin films of MnSi-type B20 metals and transition metals and their alloys.
\end{abstract}

\maketitle

\noindent
\underline{\it Introduction:}
Search for magnetic materials supporting unusual spin textures has become an important theme of research in recent years \cite{Woo2016, Yu2018, Yu2010a, Hoffmann2017, Nayak2017, Kurumaji2017, Kanazawa2012}. Presence of such textures in insulators and metals holds promise for technological applications \cite{Fert2017, Laurita2017, Wiesendanger2016}. In particular, topologically protected magnetic textures such as skyrmions, are
considered building blocks of race-track memory devices \cite{Fert2013, Nagaosa2013b, Gobel2019, Karube2018}. Presence of such spin textures in metals allows for their control using ultra-low currents. Furthermore, noncoplanar magnetic states in metals are
known to dramatically influence the spin-polarized charge transport -- a feature that can be utilized in spintronics applications \cite{Zhou2019, Kindervater2019, Zang2011, Sorn2019, Gao2018, Barcza2010, Woo2018}. There are various metallic magnets, {\it e.g.} MnSi, FeGe, Co-Zn-Mn alloys, etc., that support exotic spin textures not only in the ground state but also at higher temperatures \cite{Stishov2007, Nayak2017, Yu2011, Zhao2016, Pfleiderer2004}. Similar spin textures are also observed in thin films as well as multilayers involving transition metals \cite{Dupe2014, Pollard2017, Soumyanarayanan2017, Meyer2019}.

The key step towards designing or discovering materials with unconventional spin textures is to understand the physics of minimal microscopic models incorporating essential elementary mechanisms \cite{Farrell2014a, Chen2016, Rossler2010}. Spin Hamiltonians naturally emerge in insulators as the charge degrees of freedom become inactive and the low energy physics is determined by the spin degrees of freedom. In contrast, spin Hamiltonians in metals are phenomenologically motivated. Exceptions exist in metals that consist of a subsystem of localized magnetic moments interacting with conduction band.  
The RKKY model is a famous example in this category \cite{Ruderman1954, Kasuya1956a, Yosida1957a, Hayami2018, Bezvershenko2018}. 
Explanation of skyrmion-like spin textures relies on the presence of DM interactions \cite{Seki2012b, Seki2012c, Dzyaloshinsky1958, Moriya1960, Rossler2006}. 
However, such anisotropic terms are derived by invoking the effect of spin-orbit coupling (SOC) on Mott insulators \cite{Farrell2014a}, and should not be used for metals.
Therefore, a consistent microscopic description of exotic spin textures in metallic magnets is currently missing.

In this work, we present a closed form expression for a spin Hamiltonian for Rashba coupled double-exchange (DE) magnets. The resulting model consists of anisotropic terms resembling DM and Kitaev interactions, and it is the first example of a frustrated spin Hamiltonian for metals with nearest neighbor (nn) interactions.
After presenting the derivation, we explicitly test the validity of the pure spin model by comparing results against exact diagonalization based simulations on the starting electronic model.
The magnetic phase diagram of the new spin model is obtained via large-scale Monte Carlo simulations. The model supports, in addition to a ferromagnetic (FM) phase,
(i) single-Q (SQ) spiral states, (ii) diagonally-oriented flux (d-Flux) state, (iii) multiple-Q (MQ) states with noncoplanar skyrmion crystal (SkX) patterns, and (iv) a classical spin liquid (CSL) state characterized by diffuse ring patterns in the spin structure factor (SSF). The CSL state shows filamentary domain wall structure of remarkable similarity to the experimental data on thin films and multilayers of B20 compounds and transition metals \cite {Soumyanarayanan2017, Pollard2017, Woo2018}.
The new spin model introduced here has wide range of applicability as it originates from the FM Kondo lattice model (FKLM) -- a generic model for metals with local moments.
Some of the well known families of materials where FKLM is realized are, manganites, doped magnetic semiconductors and Heusler compounds \cite{Dagotto2002, Alvarez2002, Berciu2001, Pradhan2017, Yaouanc2020, Bombor2013, Felser2015, Sasoglu2008}. 

\noindent
\underline{\it Derivation of the spin Hamiltonian:}
Our starting point is the FKLM in the presence of Rashba SOC on a square lattice, described by the Hamiltonian,

\begin{eqnarray}
H & = & - t \sum_{\langle ij \rangle,\sigma} (c^\dagger_{i\sigma} c^{}_{j\sigma} + {\textrm H.c.}) 
+ \lambda \sum_{i} [(c^{\dagger}_{i \downarrow} c^{}_{i+x\uparrow} - c^{\dagger}_{i\uparrow} c^{}_{i+x\downarrow}) \nonumber \\
& & + \textrm{i} (c^{\dagger}_{i\downarrow} c^{}_{i+y\uparrow} + c^{\dagger}_{i\uparrow} c^{}_{i+y\downarrow}) + {\textrm H.c.}] - J_H \sum_{i} {\bf S}_i \cdot {\bf s}_i.
\label{Ham}
\end{eqnarray}
\noindent
Here, $c_{i\sigma} (c_{i\sigma}^\dagger$) annihilates (creates) an electron
at site ${i}$ with spin $\sigma$, $\langle ij \rangle$ implies that $i$ and $j$ are nn sites. $\lambda$ and $J_H$ denote the strengths of Rashba coupling and ferromagnetic Kondo (or Hund's) coupling, respectively.
$\bf{s}_i$ is the electronic spin operator at site $i$, and ${\bf S}_i$, with $|{\bf S}_i| = 1$, denotes the localized spin at that site. 
We parameterize $t = (1-\alpha) t_0$ and $\lambda = \alpha t_0$ in order to connect the weak and the strong Rashba limits, $\alpha = 0$ and $\alpha = 1$, respectively. $t_0=1$ sets the reference energy scale.

Note that coupling between localized spins ${\bf S}_i$ is mediated via the conduction electrons. In the limit of weak Kondo coupling, this leads to a modified RKKY Hamiltonian which is discussed in a recent work \cite{Okada2018}.  
To clarify the physics of the above Hamiltonian in the large $J_H$ limit, also known as the DE limit, we rewrite the Hamiltonian in a basis where the spin-quantization axes are site dependent and align with the direction of the local magnetic moment \cite{SM}. 
Since antiparallel orientations are strongly suppressed for large $J_H$, the low energy physics is determined by effectively spinless fermions with the spin quantization axis parallel to the local moments.
Projecting onto the parallel subspace, we obtain the Rashba DE (RDE) Hamiltonian,

\begin{eqnarray}
H_{{\rm RDE}} &=& \sum_{\langle ij \rangle, \gamma} [g^{\gamma}_{ij} d^{\dagger}_{ip} d^{}_{jp} + {\textrm H.c.}],
\label{Ham-DE}
\end{eqnarray}
\noindent where, $d^{}_{ip} (d^{\dagger}_{ip})$  annihilates (creates) an electron at site ${i}$ with spin parallel to the localized spin. Site $j = i + \gamma$ is the nn of site $i$ along spatial direction $\gamma = x,y$. The projected hopping $g^{\gamma}_{ij} = t^{\gamma}_{ij} + \lambda^{\gamma}_{ij}$ have contributions from the standard hopping integral $t$ and the Rashba coupling $\lambda$, and depend on the orientations of the local moments.
The two contributions to $g^{\gamma}_{ij}$ are given by,
\begin{eqnarray}
t^{\gamma}_{ij} & = & -t \big[\cos(\frac{\theta_i}{2}) \cos(\frac{\theta_j}{2}) 
+ \sin(\frac{\theta_i}{2})  \sin(\frac{\theta_j}{2})e^{-\textrm{i} (\phi_i-\phi_j)} \big],
\nonumber \\
\lambda_{{ij}}^{x} & = & \lambda \big[\sin(\frac {\theta_i}{2})  \cos(\frac {\theta_j}{2})e^{-\textrm{i} \phi_i} - \cos(\frac {\theta_i}{2})  \sin(\frac {\theta_j}{2})e^{\textrm{i} \phi_j}\big],
\nonumber \\ 
\lambda_{{ij}}^y & = & \textrm{i} \lambda \big[\sin(\frac {\theta_i}{2})  \cos(\frac {\theta_j}{2})e^{-\textrm{i} \phi_i} + \cos(\frac {\theta_i}{2})  \sin(\frac {\theta_j}{2})e^{\textrm{i} \phi_j}\big].
\end{eqnarray}
\noindent
Writing $g^{\gamma}_{ij}$ in the polar form, $g^{\gamma}_{ij} = f^{\gamma}_{ij} e^{{\rm i} h^{\gamma}_{ij}}$, and defining the ground state expectation values 
$ D^{\gamma}_{ij} = \langle [e^{{\rm i} h^{\gamma}_{ij}} d^{\dagger}_{ip} d^{}_{jp} + {\textrm H. c.}] \rangle_{gs}$ as coupling constants, we obtain the low-energy spin Hamiltonian,

\begin{eqnarray}
H_{{\rm S}} &=& -\sum_{\langle ij \rangle, \gamma} D^{\gamma}_{ij}f^{\gamma}_{ij}, \nonumber \\
\sqrt{2} f^{x}_{ij} & = &  \big[ t^2(1+{\bf S}_i \cdot {\bf S}_j)+\lambda^2(1-{\bf S}_i \cdot {\bf S}_j+2S_i^{y}S_j^{y}) \nonumber  \\
& & + 2t\lambda  \hat{y} \cdot ({\bf S}_i \times{\bf S}_j) \big]^{1/2}, \nonumber \\
\sqrt{2} f^{y}_{ij} & = &  \big[ t^2(1+{\bf S}_i \cdot {\bf S}_j)+\lambda^2(1-{\bf S}_i \cdot {\bf S}_j+2S_i^{x}S_j^{x}) \nonumber  \\
& & -2t\lambda  \hat{x} \cdot ({\bf S}_i \times{\bf S}_j) \big]^{1/2}.
\label{Ham-eff}
\end{eqnarray}

\begin{figure}[t!]
\includegraphics[width=.92 \columnwidth,angle=0,clip=true]{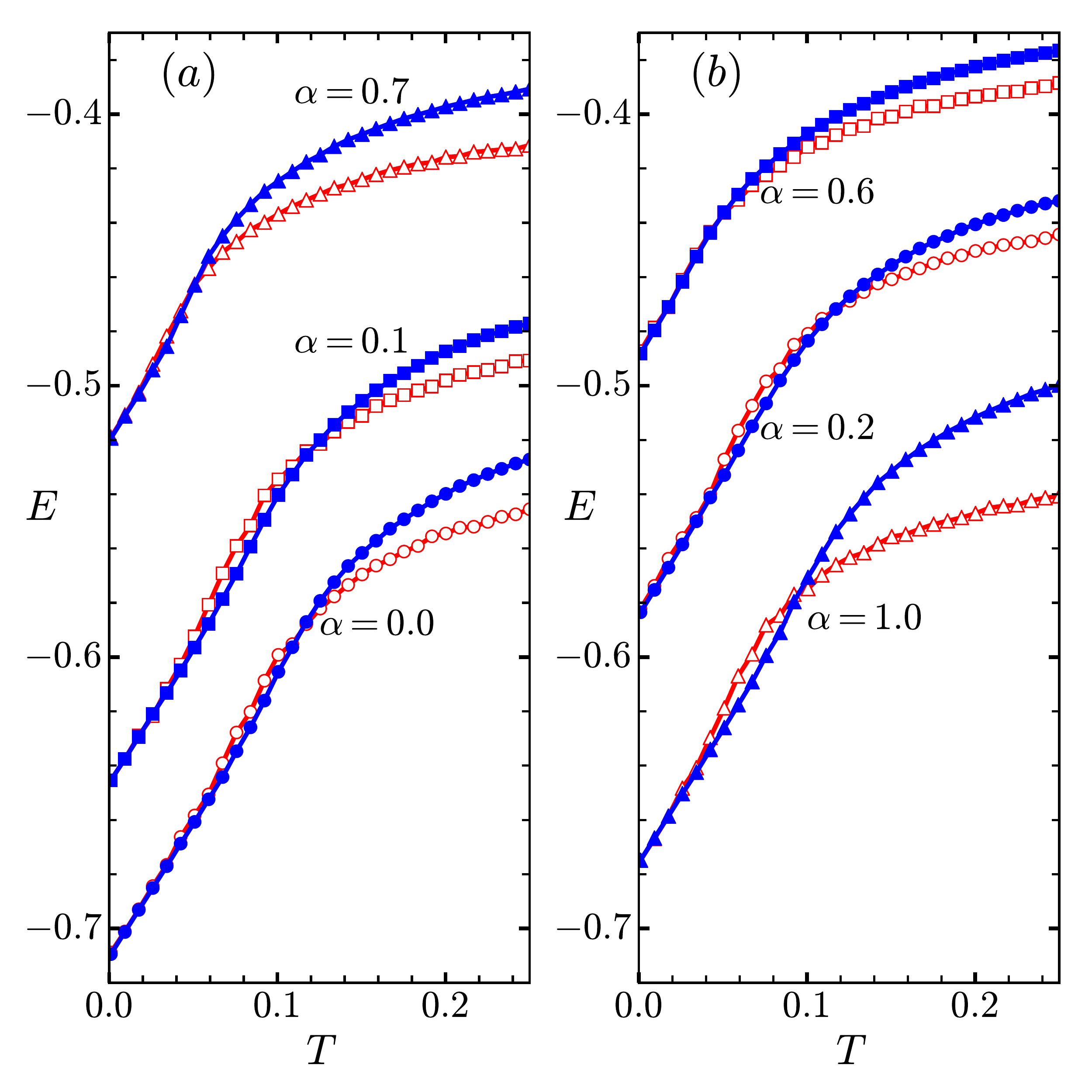}
\caption{($a$)-($b$) Temperature dependence of energy per site obtained via EDMC simulations of $H_{{\rm RDE}}$ (open symbols) and that obtained via classical Monte Carlo on $H_{{\rm S}}$ (filled symbols) for the values of $\alpha$ indicated in the panels. Simulations are carried out on $8 \times 8$ lattices.
}
\label{fig1}
\end{figure}

\noindent 
\underline{\it Comparison with the exact electronic model:}
The key question is, how well does $H_{{\rm S}}$ Eq. (\ref{Ham-eff}) describe the low energy magnetic states of the spin-fermion model $H_{{\rm RDE}}$? We directly address this by comparing energetics of the two models in the low temperature regime. Hybrid simulations combining exact diagonalization and Monte Carlo (EDMC) are carried out for $H_{{\rm RDE}}$ at electronic filling fraction of $n = 0.3$ \cite{Yunoki1998b, Dagotto2002}. Results are compared with simulations on $H_{{\rm S}}$ using $D^{\gamma}_{ij}$ as coupling constants. Energy per site $E$ is defined as statistical average $ \overline{H_{{\rm S}}}/N$ for the pure spin model, and as quantum statistical average $\overline{\langle H_{{\rm RDE}} \rangle}/N$ for the spin-fermion model, where the bar denotes the averaging over Monte Carlo steps and $N$ is the number of lattice sites.
Comparison of energy per site with varying temperature is shown for representative values of $\alpha$ (see Fig. \ref{fig1} ($a$)-($b$)). 
Ground states are correctly captured by $H_{{\rm S}}$ for all choices of $\alpha$, and the energies between $H_{{\rm RDE}}$ and $H_{{\rm S}}$ match very well in the low temperature regime. The quantitative agreement can be further improved by using simulation techniques already known for DE systems \cite{Kumar2005a, Calderon1998a}.
More importantly, we find that most of the ground states obtained in EDMC on $H_{{\rm RDE}}$ lead to values of $D^{\gamma}_{ij}$ that are independent of $ij$ \cite{SM}. 
This leads to a simplified effective spin Hamiltonian with $D^{\gamma}_{ij} \equiv D_0$ in Eq. (\ref{Ham-eff}). We will now describe the magnetic properties of this effective model using large scale Monte Carlo simulations.

\begin{figure}[t!]
\includegraphics[width=.92 \columnwidth,angle=0,clip=true]{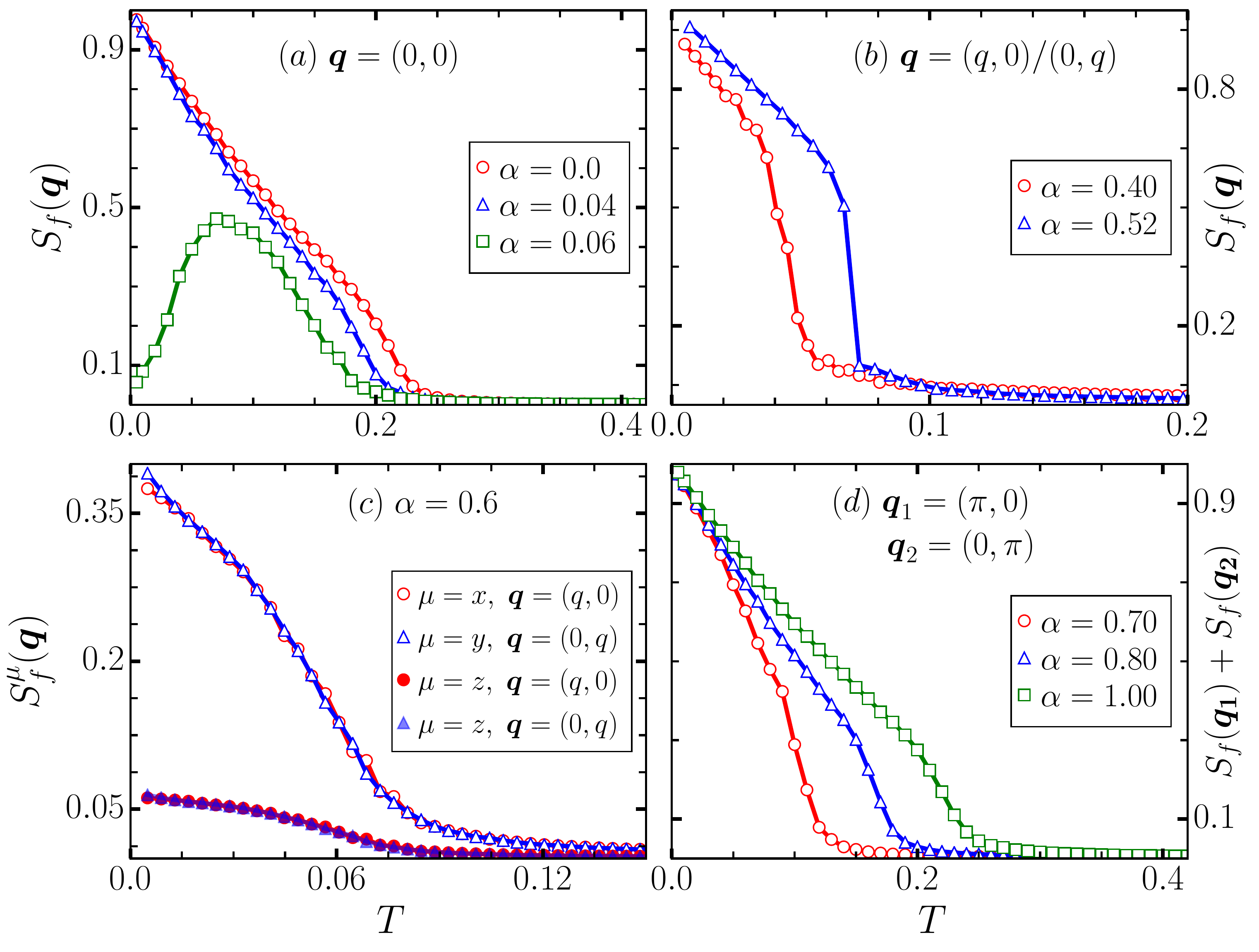}
\caption{($a$)-($d$) Temperature dependence of different components of SSF for representative values of $\alpha$. Results are obtained on $60 \times 60$ lattice.
}
\label{fig2}
\end{figure}

\noindent 
\underline{\it Magnetic phases of the new spin Hamiltonian:}
In order to investigate the magnetic phase diagram of the spin Hamiltonian Eq. (\ref{Ham-eff}) with $D^{\gamma}_{ij} \equiv D_0 = 1$, we use classical Monte Carlo simulations with the standard Metropolis algorithm. The simulations are carried out on lattice sizes varying from $N = 40^2$ to $N = 200^2$, and $\sim 5\times 10^4$ Monte Carlo steps are used for equilibration and averaging at each temperature point. The phases are characterized with the help of component resolved SSF,

\begin{eqnarray}
S^{\mu}_{f}({\bf q}) &=& \frac{1}{N^2}\sum_{ij} \overline{S^{\mu}_i S^{\mu}_j}~ e^{-{\rm i}{\bf q} \cdot ({\bf r}_i - {\bf r}_j)},
\label{SSF}
\end{eqnarray}

\noindent
where, $\mu = x, y, z$ denotes the component of the spin vector and ${\bf r}_i$ is the position vector for spin ${\bf S}_i$. The total structure factor can be computed as, 
$S_f({\bf q}) = \sum_{\mu} S^{\mu}_{f}({\bf q})$. 
Fig. \ref{fig2} shows the temperature variations of characteristic features in the SSF for different values of $\alpha$. In the small $\alpha$ regime, the ground state is FM (characterized by $S_f({\bf q})$ at ${\bf q} = (0,0)$ in Fig. \ref{fig2}($a$)) and the Curie temperature reduces with increasing $\alpha$. In the large $\alpha$ limit, d-Flux state characterized by simultaneous appearance of peaks at ${\bf q}=(\pi,0)$ and ${\bf q} = (0, \pi)$ in SSF is stabilized. The corresponding ordering temperature increases with increasing $\alpha$ (see Fig. \ref{fig2}($d$)). We find two other ordered states at intermediate values of $\alpha$: SQ spiral states with SSF peaks either at ${\bf q} = (q,0)$ or at ${\bf q} = (0,q)$ (see Fig. \ref{fig2}($b$)), and noncoplanar MQ states with all three components, $\mu = x,y,z$, contributing to total SSF at different ${\bf q}$.
For $0.06 \leq \alpha \leq 0.34$, the SSF displays a circular pattern without any prominent peaks, suggestive of a liquid-like magnetic state \cite{Tokiwa2014, Okabe2019, Nakatsuji2006}. The detailed form of SSF for these unusual phases is discussed below. 

\begin{figure}[t!]
\includegraphics[width=.92 \columnwidth,angle=0,clip=true]{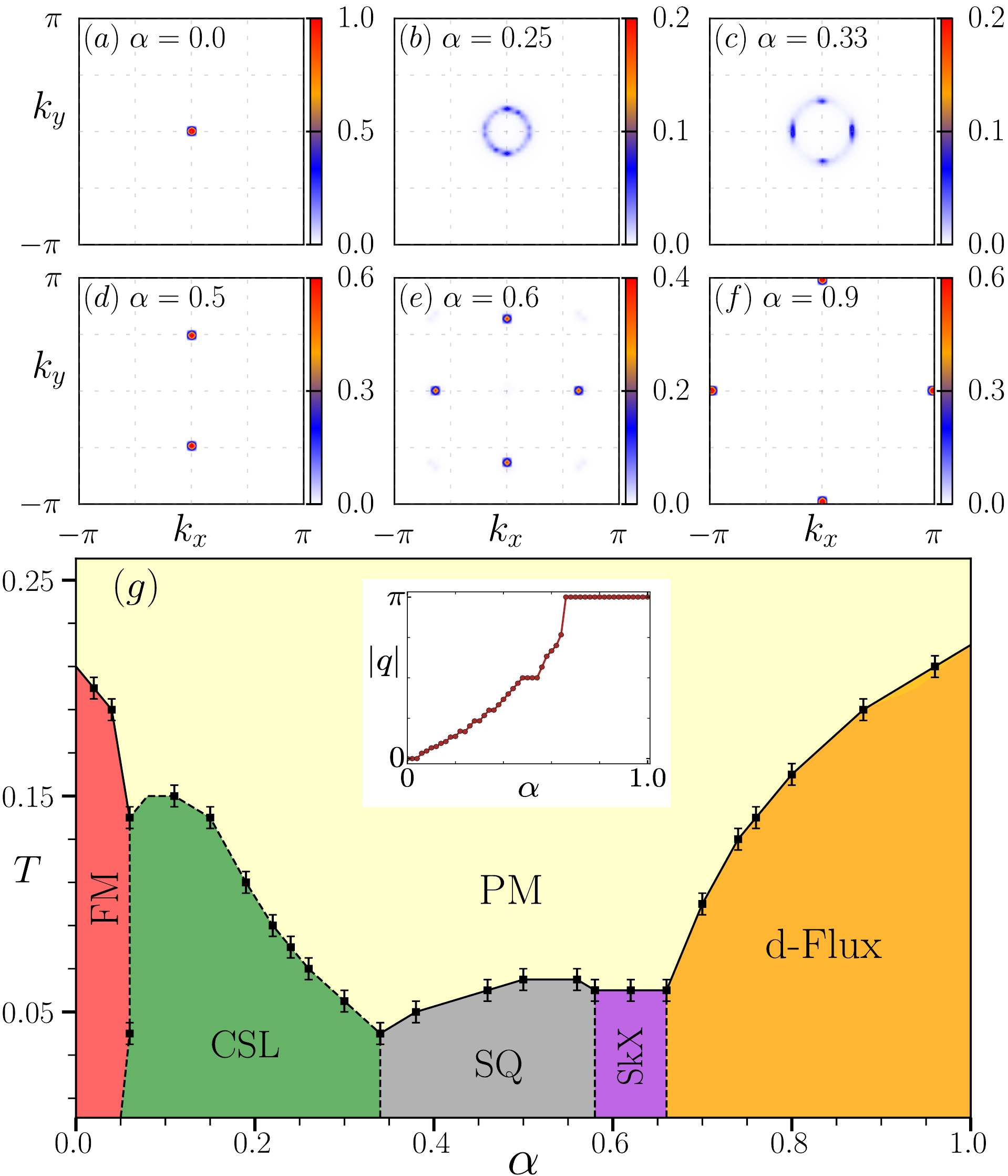}
\caption{($a$)-($f$) Color map of SSF at $T = 0.001$ for different values of $\alpha$. ($g$) Phase diagram for the new spin Hamiltonian in the $T-\alpha$ plane. The boundaries are based on the temperature dependence of the relevant components of the SSF. Inset in ($g$) shows variation in the magnitude of the relevant wave-vector with $\alpha$. 
}
\label{fig3}
\end{figure}

\begin{figure}[t!]
\includegraphics[width=.92 \columnwidth,angle=0,clip=true]{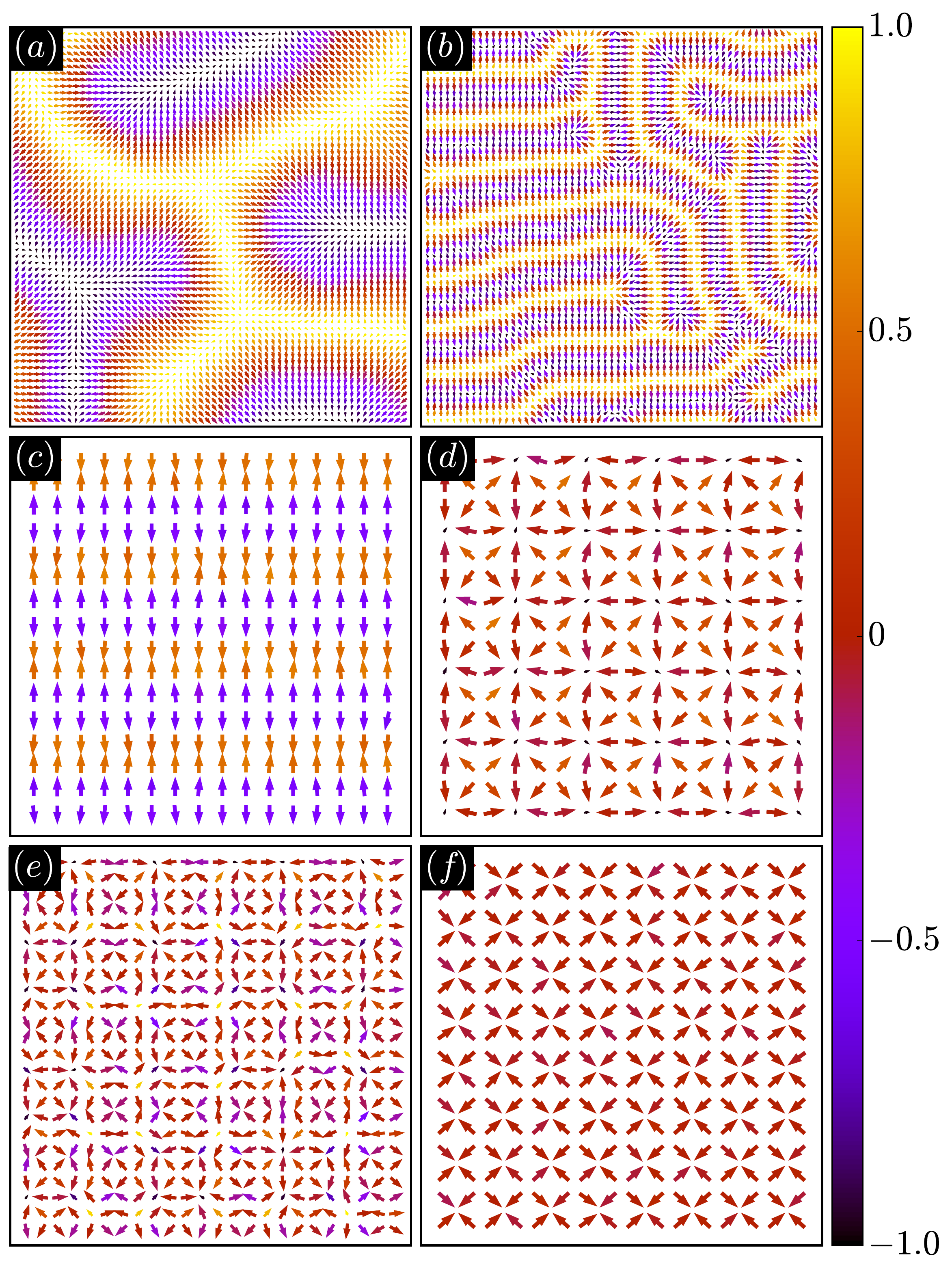}
\caption{Snapshots of spin configurations obtained at low temperature for, ($a$) $\alpha = 0.10$, ($b$) $\alpha = 0.34$,
($c$) $\alpha = 0.50$, ($d$) $\alpha = 0.60$, ($e$) $\alpha = 0.66$, and ($f$) $\alpha = 0.80$. The $x$ and $y$ components of the spins are indicated by the arrow while the $z$ component is color coded. For ($a$)-($b$) we show $60 \times 60$ lattice. For the ordered states we display for clarity only a smaller section, $16 \times 16$ for ($c$), ($d$) and ($f$) and $24 \times 24$ for ($e$), of the full lattice.
}
\label{fig4}
\end{figure}

\noindent
We summarize the simulation results in the form of a phase diagram in Fig. \ref{fig3}($g$). The ground state changes from a FM at small $\alpha$ to a d-Flux at large $\alpha$, via three non-trivial phases for intermediate values of $\alpha$. 
The evolution of the ground state SSF is displayed in Fig. \ref{fig3}($a$)-($f$). As the FM state is destabilized upon increasing $\alpha$, we do not find any ordered phase. Instead, the SSF shows a diffuse circular pattern (see Fig. \ref{fig3}($b$)) characteristic of a disordered liquid-like state. The radius of the ring increases upon increasing $\alpha$, and the intensity near the axial points, $(\pm q, 0)$ and $(0, \pm q)$, becomes relatively large (see Fig. \ref{fig3}($c$)). For $0.34 < \alpha < 0.58$, we find SQ spiral states with either horizontal or vertical FM stripes (see Fig. \ref{fig3}($d$) and Fig. \ref{fig4}($c$)). In a narrow window, $0.58 < \alpha < 0.66$, MQ noncoplanar states are stabilized. Finally the planar d-Flux state is obtained as the ground state for $\alpha > 0.66$.
Inflexion point in the temperature dependence of relevant components of SSF are used to identify the boundaries between the paramagnet (PM) and ordered phases. Note that, in case of CSL state a well defined order parameter does not exist, and dashed line indicates the temperature at which the diffuse ring pattern appears in the SSF.

We provide a clear understanding of the ground state evolution in terms of typical low temperature spin configurations in Fig. \ref{fig4}.
Upon increasing $\alpha$, the FM state is destabilized and typical configurations consist of filamentary structures of domain walls (see Fig. \ref{fig4}($a$)-($b$)).
The stability of the filamentary structures is related to an unusual degeneracy of spiral states that originates from the presence of 
mutually orthogonal directions of the two DM vectors in our spin model \cite{SM}.
The fact that domain walls can turn in arbitrary direction with negligible energy cost is responsible for the presence of the diffuse circular pattern in the SSF (see Fig. \ref{fig3} ($b$)). For larger values of $\alpha$, the width of domain walls decreases and a preference for horizontal or vertical orientations of the domain walls is found (see Fig. \ref{fig4} ($b$)). This is reflected in the appearance of arc features in SSF near the axial points (see Fig. \ref{fig3} ($c$)). For $\alpha > 0.58$ we obtain long-range ordered SQ and MQ states. The MQ states can be non-coplanar (see Fig. \ref{fig4} ($d$)-($e$)) or coplanar (see Fig. \ref{fig4} ($f$)). The noncoplanar patterns in the MQ states are identical to 
lattices of smallest skyrmions \cite{McKeever2019}. 

\noindent
\underline{\it Conclusion:} 
We have derived a new spin Hamiltonian for DE metals in the presence of Rashba SOC. Anisotropic interactions, similar to those required for stabilizing exotic spin textures, naturally arise in our model. We explicitly compare the energetics in the low temperature regime between the exact Hamiltonian and our spin model in order to prove the validity of the latter. 
Increasing the relative strength of Rashba term w.r.t. the hopping generates CSL, SQ spiral and MQ SkX states, starting from the trivial FM phase. An elegant description of this evolution emerges from the ground state degeneracy analysis.
Our spin model provides a consistent description of spin textures in itinerant magnets. In particular, the filamentary domain wall structures obtained in our simulations are in excellent agreement with the experimental observations in thin films and multilayers of transition metals \cite{Soumyanarayanan2017, Pollard2017, Woo2017, Woo2018}. Our results predict that inducing Rashba SOC in DE metals is a robust approach to generate exotic noncoplanar spin textures. 

The weak coupling approach to understand magnetism in spin orbit coupled itinerant magnets is via RKKY type effective models \cite{Okada2018}. Such models are long ranged and strongly depend on the filling fraction of the conduction band. In contrast, the form of 
our spin Hamiltonian is independent of the electronic filling fraction. Therefore, in our description, the exotic magnetic states do not originate from Fermi surface nesting features. Consequently, such states are expected without fine-tuning of electron density. This is consistent with the fact that such spin textures are experimentally observed in a variety of thin films and multilayers of transition metals. While the model is derived starting from the FKLM, at the mean-field level similar physics should hold for the Hubbard model where localized and itinerant electrons are associated with the same band \cite{Martin2008a, Pasrija2016}.
Furthermore, short-range interactions and a closed form expression are two highly desirable features of any model Hamiltonian. Therefore, in addition to its applicability in understanding magnetism of Rashba coupled itinerant systems, the new spin model should attract attention from pure statistical mechanics viewpoint.

\noindent
\underline{\it Acknowledgments:}
We acknowledge the use of computing facility at IISER Mohali.


\begin{thebibliography}{62}%
\makeatletter
\providecommand \@ifxundefined [1]{%
 \@ifx{#1\undefined}
}%
\providecommand \@ifnum [1]{%
 \ifnum #1\expandafter \@firstoftwo
 \else \expandafter \@secondoftwo
 \fi
}%
\providecommand \@ifx [1]{%
 \ifx #1\expandafter \@firstoftwo
 \else \expandafter \@secondoftwo
 \fi
}%
\providecommand \natexlab [1]{#1}%
\providecommand \enquote  [1]{``#1''}%
\providecommand \bibnamefont  [1]{#1}%
\providecommand \bibfnamefont [1]{#1}%
\providecommand \citenamefont [1]{#1}%
\providecommand \href@noop [0]{\@secondoftwo}%
\providecommand \href [0]{\begingroup \@sanitize@url \@href}%
\providecommand \@href[1]{\@@startlink{#1}\@@href}%
\providecommand \@@href[1]{\endgroup#1\@@endlink}%
\providecommand \@sanitize@url [0]{\catcode `\\12\catcode `\$12\catcode
  `\&12\catcode `\#12\catcode `\^12\catcode `\_12\catcode `\%12\relax}%
\providecommand \@@startlink[1]{}%
\providecommand \@@endlink[0]{}%
\providecommand \url  [0]{\begingroup\@sanitize@url \@url }%
\providecommand \@url [1]{\endgroup\@href {#1}{\urlprefix }}%
\providecommand \urlprefix  [0]{URL }%
\providecommand \Eprint [0]{\href }%
\providecommand \doibase [0]{http://dx.doi.org/}%
\providecommand \selectlanguage [0]{\@gobble}%
\providecommand \bibinfo  [0]{\@secondoftwo}%
\providecommand \bibfield  [0]{\@secondoftwo}%
\providecommand \translation [1]{[#1]}%
\providecommand \BibitemOpen [0]{}%
\providecommand \bibitemStop [0]{}%
\providecommand \bibitemNoStop [0]{.\EOS\space}%
\providecommand \EOS [0]{\spacefactor3000\relax}%
\providecommand \BibitemShut  [1]{\csname bibitem#1\endcsname}%
\let\auto@bib@innerbib\@empty
\bibitem [{\citenamefont {Woo}\ \emph {et~al.}(2016)\citenamefont {Woo},
  \citenamefont {Litzius}, \citenamefont {Kr{\"{u}}ger}, \citenamefont {Im},
  \citenamefont {Caretta}, \citenamefont {Richter}, \citenamefont {Mann},
  \citenamefont {Krone}, \citenamefont {Reeve}, \citenamefont {Weigand},
  \citenamefont {Agrawal}, \citenamefont {Lemesh}, \citenamefont {Mawass},
  \citenamefont {Fischer}, \citenamefont {Kl{\"{a}}ui},\ and\ \citenamefont
  {Beach}}]{Woo2016}%
  \BibitemOpen
  \bibfield  {author} {\bibinfo {author} {\bibfnamefont {S.}~\bibnamefont
  {Woo}}, \bibinfo {author} {\bibfnamefont {K.}~\bibnamefont {Litzius}},
  \bibinfo {author} {\bibfnamefont {B.}~\bibnamefont {Kr{\"{u}}ger}}, \bibinfo
  {author} {\bibfnamefont {M.~Y.}\ \bibnamefont {Im}}, \bibinfo {author}
  {\bibfnamefont {L.}~\bibnamefont {Caretta}}, \bibinfo {author} {\bibfnamefont
  {K.}~\bibnamefont {Richter}}, \bibinfo {author} {\bibfnamefont
  {M.}~\bibnamefont {Mann}}, \bibinfo {author} {\bibfnamefont {A.}~\bibnamefont
  {Krone}}, \bibinfo {author} {\bibfnamefont {R.~M.}\ \bibnamefont {Reeve}},
  \bibinfo {author} {\bibfnamefont {M.}~\bibnamefont {Weigand}}, \bibinfo
  {author} {\bibfnamefont {P.}~\bibnamefont {Agrawal}}, \bibinfo {author}
  {\bibfnamefont {I.}~\bibnamefont {Lemesh}}, \bibinfo {author} {\bibfnamefont
  {M.~A.}\ \bibnamefont {Mawass}}, \bibinfo {author} {\bibfnamefont
  {P.}~\bibnamefont {Fischer}}, \bibinfo {author} {\bibfnamefont
  {M.}~\bibnamefont {Kl{\"{a}}ui}}, \ and\ \bibinfo {author} {\bibfnamefont
  {G.~S.}\ \bibnamefont {Beach}},\ }\href {\doibase 10.1038/nmat4593}
  {\bibfield  {journal} {\bibinfo  {journal} {Nat. Mater.}\ }\textbf {\bibinfo
  {volume} {15}} (\bibinfo {year} {2016}),\ 10.1038/nmat4593},\ \Eprint
  {http://arxiv.org/abs/1502.07376} {arXiv:1502.07376} \BibitemShut {NoStop}%
\bibitem [{\citenamefont {Yu}\ \emph {et~al.}(2018)\citenamefont {Yu},
  \citenamefont {Koshibae}, \citenamefont {Tokunaga}, \citenamefont {Shibata},
  \citenamefont {Taguchi}, \citenamefont {Nagaosa},\ and\ \citenamefont
  {Tokura}}]{Yu2018}%
  \BibitemOpen
  \bibfield  {author} {\bibinfo {author} {\bibfnamefont {X.~Z.}\ \bibnamefont
  {Yu}}, \bibinfo {author} {\bibfnamefont {W.}~\bibnamefont {Koshibae}},
  \bibinfo {author} {\bibfnamefont {Y.}~\bibnamefont {Tokunaga}}, \bibinfo
  {author} {\bibfnamefont {K.}~\bibnamefont {Shibata}}, \bibinfo {author}
  {\bibfnamefont {Y.}~\bibnamefont {Taguchi}}, \bibinfo {author} {\bibfnamefont
  {N.}~\bibnamefont {Nagaosa}}, \ and\ \bibinfo {author} {\bibfnamefont
  {Y.}~\bibnamefont {Tokura}},\ }\href {\doibase 10.1038/s41586-018-0745-3}
  {\bibfield  {journal} {\bibinfo  {journal} {Nature}\ }\textbf {\bibinfo
  {volume} {564}},\ \bibinfo {pages} {95} (\bibinfo {year} {2018})}\BibitemShut
  {NoStop}%
\bibitem [{\citenamefont {Yu}\ \emph {et~al.}(2010)\citenamefont {Yu},
  \citenamefont {Onose}, \citenamefont {Kanazawa}, \citenamefont {Park},
  \citenamefont {Han}, \citenamefont {Matsui}, \citenamefont {Nagaosa},\ and\
  \citenamefont {Tokura}}]{Yu2010a}%
  \BibitemOpen
  \bibfield  {author} {\bibinfo {author} {\bibfnamefont {X.~Z.}\ \bibnamefont
  {Yu}}, \bibinfo {author} {\bibfnamefont {Y.}~\bibnamefont {Onose}}, \bibinfo
  {author} {\bibfnamefont {N.}~\bibnamefont {Kanazawa}}, \bibinfo {author}
  {\bibfnamefont {J.~H.}\ \bibnamefont {Park}}, \bibinfo {author}
  {\bibfnamefont {J.~H.}\ \bibnamefont {Han}}, \bibinfo {author} {\bibfnamefont
  {Y.}~\bibnamefont {Matsui}}, \bibinfo {author} {\bibfnamefont
  {N.}~\bibnamefont {Nagaosa}}, \ and\ \bibinfo {author} {\bibfnamefont
  {Y.}~\bibnamefont {Tokura}},\ }\href {\doibase 10.1038/nature09124}
  {\bibfield  {journal} {\bibinfo  {journal} {Nature}\ }\textbf {\bibinfo
  {volume} {465}},\ \bibinfo {pages} {901} (\bibinfo {year}
  {2010})}\BibitemShut {NoStop}%
\bibitem [{\citenamefont {Hoffmann}\ \emph {et~al.}(2017)\citenamefont
  {Hoffmann}, \citenamefont {Zimmermann}, \citenamefont {M{\"{u}}ller},
  \citenamefont {Sch{\"{u}}rhoff}, \citenamefont {Kiselev}, \citenamefont
  {Melcher},\ and\ \citenamefont {Bl{\"{u}}gel}}]{Hoffmann2017}%
  \BibitemOpen
  \bibfield  {author} {\bibinfo {author} {\bibfnamefont {M.}~\bibnamefont
  {Hoffmann}}, \bibinfo {author} {\bibfnamefont {B.}~\bibnamefont
  {Zimmermann}}, \bibinfo {author} {\bibfnamefont {G.~P.}\ \bibnamefont
  {M{\"{u}}ller}}, \bibinfo {author} {\bibfnamefont {D.}~\bibnamefont
  {Sch{\"{u}}rhoff}}, \bibinfo {author} {\bibfnamefont {N.~S.}\ \bibnamefont
  {Kiselev}}, \bibinfo {author} {\bibfnamefont {C.}~\bibnamefont {Melcher}}, \
  and\ \bibinfo {author} {\bibfnamefont {S.}~\bibnamefont {Bl{\"{u}}gel}},\
  }\href {\doibase 10.1038/s41467-017-00313-0} {\bibfield  {journal} {\bibinfo
  {journal} {Nat. Commun.}\ }\textbf {\bibinfo {volume} {8}},\ \bibinfo {pages}
  {308} (\bibinfo {year} {2017})}\BibitemShut {NoStop}%
\bibitem [{\citenamefont {Nayak}\ \emph {et~al.}(2017)\citenamefont {Nayak},
  \citenamefont {Kumar}, \citenamefont {Ma}, \citenamefont {Werner},
  \citenamefont {Pippel}, \citenamefont {Sahoo}, \citenamefont {Damay},
  \citenamefont {R{\"{o}}{\ss}ler}, \citenamefont {Felser},\ and\ \citenamefont
  {Parkin}}]{Nayak2017}%
  \BibitemOpen
  \bibfield  {author} {\bibinfo {author} {\bibfnamefont {A.~K.}\ \bibnamefont
  {Nayak}}, \bibinfo {author} {\bibfnamefont {V.}~\bibnamefont {Kumar}},
  \bibinfo {author} {\bibfnamefont {T.}~\bibnamefont {Ma}}, \bibinfo {author}
  {\bibfnamefont {P.}~\bibnamefont {Werner}}, \bibinfo {author} {\bibfnamefont
  {E.}~\bibnamefont {Pippel}}, \bibinfo {author} {\bibfnamefont
  {R.}~\bibnamefont {Sahoo}}, \bibinfo {author} {\bibfnamefont
  {F.}~\bibnamefont {Damay}}, \bibinfo {author} {\bibfnamefont {U.~K.}\
  \bibnamefont {R{\"{o}}{\ss}ler}}, \bibinfo {author} {\bibfnamefont
  {C.}~\bibnamefont {Felser}}, \ and\ \bibinfo {author} {\bibfnamefont
  {S.~S.~P.}\ \bibnamefont {Parkin}},\ }\href {\doibase 10.1038/nature23466}
  {\bibfield  {journal} {\bibinfo  {journal} {Nature}\ }\textbf {\bibinfo
  {volume} {548}},\ \bibinfo {pages} {561} (\bibinfo {year}
  {2017})}\BibitemShut {NoStop}%
\bibitem [{\citenamefont {Kurumaji}\ \emph {et~al.}(2017)\citenamefont
  {Kurumaji}, \citenamefont {Nakajima}, \citenamefont {Ukleev}, \citenamefont
  {Feoktystov}, \citenamefont {Arima}, \citenamefont {Kakurai},\ and\
  \citenamefont {Tokura}}]{Kurumaji2017}%
  \BibitemOpen
  \bibfield  {author} {\bibinfo {author} {\bibfnamefont {T.}~\bibnamefont
  {Kurumaji}}, \bibinfo {author} {\bibfnamefont {T.}~\bibnamefont {Nakajima}},
  \bibinfo {author} {\bibfnamefont {V.}~\bibnamefont {Ukleev}}, \bibinfo
  {author} {\bibfnamefont {A.}~\bibnamefont {Feoktystov}}, \bibinfo {author}
  {\bibfnamefont {T.~H.}\ \bibnamefont {Arima}}, \bibinfo {author}
  {\bibfnamefont {K.}~\bibnamefont {Kakurai}}, \ and\ \bibinfo {author}
  {\bibfnamefont {Y.}~\bibnamefont {Tokura}},\ }\href {\doibase
  10.1103/PhysRevLett.119.237201} {\bibfield  {journal} {\bibinfo  {journal}
  {Phys. Rev. Lett.}\ }\textbf {\bibinfo {volume} {119}} (\bibinfo {year}
  {2017}),\ 10.1103/PhysRevLett.119.237201},\ \Eprint
  {http://arxiv.org/abs/1710.04000} {arXiv:1710.04000} \BibitemShut {NoStop}%
\bibitem [{\citenamefont {Kanazawa}\ \emph {et~al.}(2012)\citenamefont
  {Kanazawa}, \citenamefont {Kim}, \citenamefont {Inosov}, \citenamefont
  {White}, \citenamefont {Egetenmeyer}, \citenamefont {Gavilano}, \citenamefont
  {Ishiwata}, \citenamefont {Onose}, \citenamefont {Arima}, \citenamefont
  {Keimer},\ and\ \citenamefont {Tokura}}]{Kanazawa2012}%
  \BibitemOpen
  \bibfield  {author} {\bibinfo {author} {\bibfnamefont {N.}~\bibnamefont
  {Kanazawa}}, \bibinfo {author} {\bibfnamefont {J.-H.}\ \bibnamefont {Kim}},
  \bibinfo {author} {\bibfnamefont {D.~S.}\ \bibnamefont {Inosov}}, \bibinfo
  {author} {\bibfnamefont {J.~S.}\ \bibnamefont {White}}, \bibinfo {author}
  {\bibfnamefont {N.}~\bibnamefont {Egetenmeyer}}, \bibinfo {author}
  {\bibfnamefont {J.~L.}\ \bibnamefont {Gavilano}}, \bibinfo {author}
  {\bibfnamefont {S.}~\bibnamefont {Ishiwata}}, \bibinfo {author}
  {\bibfnamefont {Y.}~\bibnamefont {Onose}}, \bibinfo {author} {\bibfnamefont
  {T.}~\bibnamefont {Arima}}, \bibinfo {author} {\bibfnamefont
  {B.}~\bibnamefont {Keimer}}, \ and\ \bibinfo {author} {\bibfnamefont
  {Y.}~\bibnamefont {Tokura}},\ }\href {\doibase 10.1103/PhysRevB.86.134425}
  {\bibfield  {journal} {\bibinfo  {journal} {Phys. Rev. B}\ }\textbf {\bibinfo
  {volume} {86}},\ \bibinfo {pages} {134425} (\bibinfo {year}
  {2012})}\BibitemShut {NoStop}%
\bibitem [{\citenamefont {Fert}\ \emph {et~al.}(2017)\citenamefont {Fert},
  \citenamefont {Reyren},\ and\ \citenamefont {Cros}}]{Fert2017}%
  \BibitemOpen
  \bibfield  {author} {\bibinfo {author} {\bibfnamefont {A.}~\bibnamefont
  {Fert}}, \bibinfo {author} {\bibfnamefont {N.}~\bibnamefont {Reyren}}, \ and\
  \bibinfo {author} {\bibfnamefont {V.}~\bibnamefont {Cros}},\ }\href {\doibase
  10.1038/natrevmats.2017.31} {\bibfield  {journal} {\bibinfo  {journal} {Nat.
  Rev. Mater.}\ }\textbf {\bibinfo {volume} {2}},\ \bibinfo {pages} {17031}
  (\bibinfo {year} {2017})}\BibitemShut {NoStop}%
\bibitem [{\citenamefont {Laurita}\ \emph {et~al.}(2017)\citenamefont
  {Laurita}, \citenamefont {Marcus}, \citenamefont {Trump}, \citenamefont
  {Kindervater}, \citenamefont {Stone}, \citenamefont {McQueen}, \citenamefont
  {Broholm},\ and\ \citenamefont {Armitage}}]{Laurita2017}%
  \BibitemOpen
  \bibfield  {author} {\bibinfo {author} {\bibfnamefont {N.~J.}\ \bibnamefont
  {Laurita}}, \bibinfo {author} {\bibfnamefont {G.~G.}\ \bibnamefont {Marcus}},
  \bibinfo {author} {\bibfnamefont {B.~A.}\ \bibnamefont {Trump}}, \bibinfo
  {author} {\bibfnamefont {J.}~\bibnamefont {Kindervater}}, \bibinfo {author}
  {\bibfnamefont {M.~B.}\ \bibnamefont {Stone}}, \bibinfo {author}
  {\bibfnamefont {T.~M.}\ \bibnamefont {McQueen}}, \bibinfo {author}
  {\bibfnamefont {C.~L.}\ \bibnamefont {Broholm}}, \ and\ \bibinfo {author}
  {\bibfnamefont {N.~P.}\ \bibnamefont {Armitage}},\ }\href {\doibase
  10.1103/PhysRevB.95.235155} {\bibfield  {journal} {\bibinfo  {journal} {Phys.
  Rev. B}\ }\textbf {\bibinfo {volume} {95}},\ \bibinfo {pages} {235155}
  (\bibinfo {year} {2017})}\BibitemShut {NoStop}%
\bibitem [{\citenamefont {Wiesendanger}(2016)}]{Wiesendanger2016}%
  \BibitemOpen
  \bibfield  {author} {\bibinfo {author} {\bibfnamefont {R.}~\bibnamefont
  {Wiesendanger}},\ }\href {\doibase 10.1038/natrevmats.2016.44} {\bibfield
  {journal} {\bibinfo  {journal} {Nat. Rev. Mater.}\ }\textbf {\bibinfo
  {volume} {1}},\ \bibinfo {pages} {16044} (\bibinfo {year}
  {2016})}\BibitemShut {NoStop}%
\bibitem [{\citenamefont {Fert}\ \emph {et~al.}(2013)\citenamefont {Fert},
  \citenamefont {Cros},\ and\ \citenamefont {Sampaio}}]{Fert2013}%
  \BibitemOpen
  \bibfield  {author} {\bibinfo {author} {\bibfnamefont {A.}~\bibnamefont
  {Fert}}, \bibinfo {author} {\bibfnamefont {V.}~\bibnamefont {Cros}}, \ and\
  \bibinfo {author} {\bibfnamefont {J.}~\bibnamefont {Sampaio}},\ }\href
  {\doibase 10.1038/nnano.2013.29} {\bibfield  {journal} {\bibinfo  {journal}
  {Nat. Nanotechnol.}\ }\textbf {\bibinfo {volume} {8}},\ \bibinfo {pages}
  {152} (\bibinfo {year} {2013})}\BibitemShut {NoStop}%
\bibitem [{\citenamefont {Nagaosa}\ and\ \citenamefont
  {Tokura}(2013)}]{Nagaosa2013b}%
  \BibitemOpen
  \bibfield  {author} {\bibinfo {author} {\bibfnamefont {N.}~\bibnamefont
  {Nagaosa}}\ and\ \bibinfo {author} {\bibfnamefont {Y.}~\bibnamefont
  {Tokura}},\ }\href {\doibase 10.1038/nnano.2013.243} {\bibfield  {journal}
  {\bibinfo  {journal} {Nat. Nanotechnol.}\ }\textbf {\bibinfo {volume} {8}},\
  \bibinfo {pages} {899} (\bibinfo {year} {2013})}\BibitemShut {NoStop}%
\bibitem [{\citenamefont {G{\"{o}}bel}\ \emph {et~al.}(2019)\citenamefont
  {G{\"{o}}bel}, \citenamefont {Mook}, \citenamefont {Henk},\ and\
  \citenamefont {Mertig}}]{Gobel2019}%
  \BibitemOpen
  \bibfield  {author} {\bibinfo {author} {\bibfnamefont {B.}~\bibnamefont
  {G{\"{o}}bel}}, \bibinfo {author} {\bibfnamefont {A.}~\bibnamefont {Mook}},
  \bibinfo {author} {\bibfnamefont {J.}~\bibnamefont {Henk}}, \ and\ \bibinfo
  {author} {\bibfnamefont {I.}~\bibnamefont {Mertig}},\ }\href {\doibase
  10.1103/PhysRevB.99.020405} {\bibfield  {journal} {\bibinfo  {journal} {Phys.
  Rev. B}\ }\textbf {\bibinfo {volume} {99}},\ \bibinfo {pages} {020405}
  (\bibinfo {year} {2019})}\BibitemShut {NoStop}%
\bibitem [{\citenamefont {Karube}\ \emph {et~al.}(2018)\citenamefont {Karube},
  \citenamefont {Shibata}, \citenamefont {White}, \citenamefont {Koretsune},
  \citenamefont {Yu}, \citenamefont {Tokunaga}, \citenamefont {R{\o}nnow},
  \citenamefont {Arita}, \citenamefont {Arima}, \citenamefont {Tokura},\ and\
  \citenamefont {Taguchi}}]{Karube2018}%
  \BibitemOpen
  \bibfield  {author} {\bibinfo {author} {\bibfnamefont {K.}~\bibnamefont
  {Karube}}, \bibinfo {author} {\bibfnamefont {K.}~\bibnamefont {Shibata}},
  \bibinfo {author} {\bibfnamefont {J.~S.}\ \bibnamefont {White}}, \bibinfo
  {author} {\bibfnamefont {T.}~\bibnamefont {Koretsune}}, \bibinfo {author}
  {\bibfnamefont {X.~Z.}\ \bibnamefont {Yu}}, \bibinfo {author} {\bibfnamefont
  {Y.}~\bibnamefont {Tokunaga}}, \bibinfo {author} {\bibfnamefont {H.~M.}\
  \bibnamefont {R{\o}nnow}}, \bibinfo {author} {\bibfnamefont {R.}~\bibnamefont
  {Arita}}, \bibinfo {author} {\bibfnamefont {T.}~\bibnamefont {Arima}},
  \bibinfo {author} {\bibfnamefont {Y.}~\bibnamefont {Tokura}}, \ and\ \bibinfo
  {author} {\bibfnamefont {Y.}~\bibnamefont {Taguchi}},\ }\href {\doibase
  10.1103/PhysRevB.98.155120} {\bibfield  {journal} {\bibinfo  {journal} {Phys.
  Rev. B}\ }\textbf {\bibinfo {volume} {98}},\ \bibinfo {pages} {155120}
  (\bibinfo {year} {2018})}\BibitemShut {NoStop}%
\bibitem [{\citenamefont {Zhou}\ \emph {et~al.}(2019)\citenamefont {Zhou},
  \citenamefont {Mohanta}, \citenamefont {Han}, \citenamefont {Matos-Abiague},\
  and\ \citenamefont {{\v{Z}}uti{\'{c}}}}]{Zhou2019}%
  \BibitemOpen
  \bibfield  {author} {\bibinfo {author} {\bibfnamefont {T.}~\bibnamefont
  {Zhou}}, \bibinfo {author} {\bibfnamefont {N.}~\bibnamefont {Mohanta}},
  \bibinfo {author} {\bibfnamefont {J.~E.}\ \bibnamefont {Han}}, \bibinfo
  {author} {\bibfnamefont {A.}~\bibnamefont {Matos-Abiague}}, \ and\ \bibinfo
  {author} {\bibfnamefont {I.}~\bibnamefont {{\v{Z}}uti{\'{c}}}},\ }\href
  {\doibase 10.1103/PhysRevB.99.134505} {\bibfield  {journal} {\bibinfo
  {journal} {Phys. Rev. B}\ }\textbf {\bibinfo {volume} {99}},\ \bibinfo
  {pages} {134505} (\bibinfo {year} {2019})}\BibitemShut {NoStop}%
\bibitem [{\citenamefont {Kindervater}\ \emph {et~al.}(2019)\citenamefont
  {Kindervater}, \citenamefont {Stasinopoulos}, \citenamefont {Bauer},
  \citenamefont {Haslbeck}, \citenamefont {Rucker}, \citenamefont {Chacon},
  \citenamefont {M{\"{u}}hlbauer}, \citenamefont {Franz}, \citenamefont
  {Garst}, \citenamefont {Grundler},\ and\ \citenamefont
  {Pfleiderer}}]{Kindervater2019}%
  \BibitemOpen
  \bibfield  {author} {\bibinfo {author} {\bibfnamefont {J.}~\bibnamefont
  {Kindervater}}, \bibinfo {author} {\bibfnamefont {I.}~\bibnamefont
  {Stasinopoulos}}, \bibinfo {author} {\bibfnamefont {A.}~\bibnamefont
  {Bauer}}, \bibinfo {author} {\bibfnamefont {F.~X.}\ \bibnamefont {Haslbeck}},
  \bibinfo {author} {\bibfnamefont {F.}~\bibnamefont {Rucker}}, \bibinfo
  {author} {\bibfnamefont {A.}~\bibnamefont {Chacon}}, \bibinfo {author}
  {\bibfnamefont {S.}~\bibnamefont {M{\"{u}}hlbauer}}, \bibinfo {author}
  {\bibfnamefont {C.}~\bibnamefont {Franz}}, \bibinfo {author} {\bibfnamefont
  {M.}~\bibnamefont {Garst}}, \bibinfo {author} {\bibfnamefont
  {D.}~\bibnamefont {Grundler}}, \ and\ \bibinfo {author} {\bibfnamefont
  {C.}~\bibnamefont {Pfleiderer}},\ }\href {\doibase 10.1103/PhysRevX.9.041059}
  {\bibfield  {journal} {\bibinfo  {journal} {Phys. Rev. X}\ }\textbf {\bibinfo
  {volume} {9}},\ \bibinfo {pages} {041059} (\bibinfo {year}
  {2019})}\BibitemShut {NoStop}%
\bibitem [{\citenamefont {Zang}\ \emph {et~al.}(2011)\citenamefont {Zang},
  \citenamefont {Mostovoy}, \citenamefont {Han},\ and\ \citenamefont
  {Nagaosa}}]{Zang2011}%
  \BibitemOpen
  \bibfield  {author} {\bibinfo {author} {\bibfnamefont {J.}~\bibnamefont
  {Zang}}, \bibinfo {author} {\bibfnamefont {M.}~\bibnamefont {Mostovoy}},
  \bibinfo {author} {\bibfnamefont {J.~H.}\ \bibnamefont {Han}}, \ and\
  \bibinfo {author} {\bibfnamefont {N.}~\bibnamefont {Nagaosa}},\ }\href
  {\doibase 10.1103/PhysRevLett.107.136804} {\bibfield  {journal} {\bibinfo
  {journal} {Phys. Rev. Lett.}\ }\textbf {\bibinfo {volume} {107}},\ \bibinfo
  {pages} {136804} (\bibinfo {year} {2011})}\BibitemShut {NoStop}%
\bibitem [{\citenamefont {Sorn}\ \emph {et~al.}(2019)\citenamefont {Sorn},
  \citenamefont {Divic},\ and\ \citenamefont {Paramekanti}}]{Sorn2019}%
  \BibitemOpen
  \bibfield  {author} {\bibinfo {author} {\bibfnamefont {S.}~\bibnamefont
  {Sorn}}, \bibinfo {author} {\bibfnamefont {S.}~\bibnamefont {Divic}}, \ and\
  \bibinfo {author} {\bibfnamefont {A.}~\bibnamefont {Paramekanti}},\ }\href
  {\doibase 10.1103/PhysRevB.100.174411} {\bibfield  {journal} {\bibinfo
  {journal} {Phys. Rev. B}\ }\textbf {\bibinfo {volume} {100}},\ \bibinfo
  {pages} {174411} (\bibinfo {year} {2019})}\BibitemShut {NoStop}%
\bibitem [{\citenamefont {Gao}\ \emph {et~al.}(2018)\citenamefont {Gao},
  \citenamefont {Qaiumzadeh}, \citenamefont {An}, \citenamefont {Musha},
  \citenamefont {Kageyama}, \citenamefont {Shi},\ and\ \citenamefont
  {Ando}}]{Gao2018}%
  \BibitemOpen
  \bibfield  {author} {\bibinfo {author} {\bibfnamefont {T.}~\bibnamefont
  {Gao}}, \bibinfo {author} {\bibfnamefont {A.}~\bibnamefont {Qaiumzadeh}},
  \bibinfo {author} {\bibfnamefont {H.}~\bibnamefont {An}}, \bibinfo {author}
  {\bibfnamefont {A.}~\bibnamefont {Musha}}, \bibinfo {author} {\bibfnamefont
  {Y.}~\bibnamefont {Kageyama}}, \bibinfo {author} {\bibfnamefont
  {J.}~\bibnamefont {Shi}}, \ and\ \bibinfo {author} {\bibfnamefont
  {K.}~\bibnamefont {Ando}},\ }\href {\doibase 10.1103/PhysRevLett.121.017202}
  {\bibfield  {journal} {\bibinfo  {journal} {Phys. Rev. Lett.}\ }\textbf
  {\bibinfo {volume} {121}},\ \bibinfo {pages} {017202} (\bibinfo {year}
  {2018})}\BibitemShut {NoStop}%
\bibitem [{\citenamefont {Barcza}\ \emph {et~al.}(2010)\citenamefont {Barcza},
  \citenamefont {Gercsi}, \citenamefont {Knight},\ and\ \citenamefont
  {Sandeman}}]{Barcza2010}%
  \BibitemOpen
  \bibfield  {author} {\bibinfo {author} {\bibfnamefont {A.}~\bibnamefont
  {Barcza}}, \bibinfo {author} {\bibfnamefont {Z.}~\bibnamefont {Gercsi}},
  \bibinfo {author} {\bibfnamefont {K.~S.}\ \bibnamefont {Knight}}, \ and\
  \bibinfo {author} {\bibfnamefont {K.~G.}\ \bibnamefont {Sandeman}},\ }\href
  {\doibase 10.1103/PhysRevLett.104.247202} {\bibfield  {journal} {\bibinfo
  {journal} {Phys. Rev. Lett.}\ }\textbf {\bibinfo {volume} {104}},\ \bibinfo
  {pages} {247202} (\bibinfo {year} {2010})}\BibitemShut {NoStop}%
\bibitem [{\citenamefont {Woo}\ \emph {et~al.}(2018)\citenamefont {Woo},
  \citenamefont {Song}, \citenamefont {Zhang}, \citenamefont {Zhou},
  \citenamefont {Ezawa}, \citenamefont {Liu}, \citenamefont {Finizio},
  \citenamefont {Raabe}, \citenamefont {Lee}, \citenamefont {Kim},
  \citenamefont {Park}, \citenamefont {Kim}, \citenamefont {Kim}, \citenamefont
  {Lee}, \citenamefont {Lee}, \citenamefont {Choi}, \citenamefont {Min},
  \citenamefont {Koo},\ and\ \citenamefont {Chang}}]{Woo2018}%
  \BibitemOpen
  \bibfield  {author} {\bibinfo {author} {\bibfnamefont {S.}~\bibnamefont
  {Woo}}, \bibinfo {author} {\bibfnamefont {K.~M.}\ \bibnamefont {Song}},
  \bibinfo {author} {\bibfnamefont {X.}~\bibnamefont {Zhang}}, \bibinfo
  {author} {\bibfnamefont {Y.}~\bibnamefont {Zhou}}, \bibinfo {author}
  {\bibfnamefont {M.}~\bibnamefont {Ezawa}}, \bibinfo {author} {\bibfnamefont
  {X.}~\bibnamefont {Liu}}, \bibinfo {author} {\bibfnamefont {S.}~\bibnamefont
  {Finizio}}, \bibinfo {author} {\bibfnamefont {J.}~\bibnamefont {Raabe}},
  \bibinfo {author} {\bibfnamefont {N.~J.}\ \bibnamefont {Lee}}, \bibinfo
  {author} {\bibfnamefont {S.-I.}\ \bibnamefont {Kim}}, \bibinfo {author}
  {\bibfnamefont {S.-Y.}\ \bibnamefont {Park}}, \bibinfo {author}
  {\bibfnamefont {Y.}~\bibnamefont {Kim}}, \bibinfo {author} {\bibfnamefont
  {J.-Y.}\ \bibnamefont {Kim}}, \bibinfo {author} {\bibfnamefont
  {D.}~\bibnamefont {Lee}}, \bibinfo {author} {\bibfnamefont {O.}~\bibnamefont
  {Lee}}, \bibinfo {author} {\bibfnamefont {J.~W.}\ \bibnamefont {Choi}},
  \bibinfo {author} {\bibfnamefont {B.-C.}\ \bibnamefont {Min}}, \bibinfo
  {author} {\bibfnamefont {H.~C.}\ \bibnamefont {Koo}}, \ and\ \bibinfo
  {author} {\bibfnamefont {J.}~\bibnamefont {Chang}},\ }\href {\doibase
  10.1038/s41467-018-03378-7} {\bibfield  {journal} {\bibinfo  {journal} {Nat.
  Commun.}\ }\textbf {\bibinfo {volume} {9}},\ \bibinfo {pages} {959} (\bibinfo
  {year} {2018})}\BibitemShut {NoStop}%
\bibitem [{\citenamefont {Stishov}\ \emph {et~al.}(2007)\citenamefont
  {Stishov}, \citenamefont {Petrova}, \citenamefont {Khasanov}, \citenamefont
  {Panova}, \citenamefont {Shikov}, \citenamefont {Lashley}, \citenamefont
  {Wu},\ and\ \citenamefont {Lograsso}}]{Stishov2007}%
  \BibitemOpen
  \bibfield  {author} {\bibinfo {author} {\bibfnamefont {S.~M.}\ \bibnamefont
  {Stishov}}, \bibinfo {author} {\bibfnamefont {A.~E.}\ \bibnamefont
  {Petrova}}, \bibinfo {author} {\bibfnamefont {S.}~\bibnamefont {Khasanov}},
  \bibinfo {author} {\bibfnamefont {G.~K.}\ \bibnamefont {Panova}}, \bibinfo
  {author} {\bibfnamefont {A.~A.}\ \bibnamefont {Shikov}}, \bibinfo {author}
  {\bibfnamefont {J.~C.}\ \bibnamefont {Lashley}}, \bibinfo {author}
  {\bibfnamefont {D.}~\bibnamefont {Wu}}, \ and\ \bibinfo {author}
  {\bibfnamefont {T.~A.}\ \bibnamefont {Lograsso}},\ }\href {\doibase
  10.1103/PhysRevB.76.052405} {\bibfield  {journal} {\bibinfo  {journal} {Phys.
  Rev. B}\ }\textbf {\bibinfo {volume} {76}},\ \bibinfo {pages} {052405}
  (\bibinfo {year} {2007})}\BibitemShut {NoStop}%
\bibitem [{\citenamefont {Yu}\ \emph {et~al.}(2011)\citenamefont {Yu},
  \citenamefont {Kanazawa}, \citenamefont {Onose}, \citenamefont {Kimoto},
  \citenamefont {Zhang}, \citenamefont {Ishiwata}, \citenamefont {Matsui},\
  and\ \citenamefont {Tokura}}]{Yu2011}%
  \BibitemOpen
  \bibfield  {author} {\bibinfo {author} {\bibfnamefont {X.~Z.}\ \bibnamefont
  {Yu}}, \bibinfo {author} {\bibfnamefont {N.}~\bibnamefont {Kanazawa}},
  \bibinfo {author} {\bibfnamefont {Y.}~\bibnamefont {Onose}}, \bibinfo
  {author} {\bibfnamefont {K.}~\bibnamefont {Kimoto}}, \bibinfo {author}
  {\bibfnamefont {W.~Z.}\ \bibnamefont {Zhang}}, \bibinfo {author}
  {\bibfnamefont {S.}~\bibnamefont {Ishiwata}}, \bibinfo {author}
  {\bibfnamefont {Y.}~\bibnamefont {Matsui}}, \ and\ \bibinfo {author}
  {\bibfnamefont {Y.}~\bibnamefont {Tokura}},\ }\href {\doibase
  10.1038/nmat2916} {\bibfield  {journal} {\bibinfo  {journal} {Nat. Mater.}\
  }\textbf {\bibinfo {volume} {10}},\ \bibinfo {pages} {106} (\bibinfo {year}
  {2011})}\BibitemShut {NoStop}%
\bibitem [{\citenamefont {Zhao}\ \emph {et~al.}(2016)\citenamefont {Zhao},
  \citenamefont {Jin}, \citenamefont {Wang}, \citenamefont {Du}, \citenamefont
  {Zang}, \citenamefont {Tian}, \citenamefont {Che},\ and\ \citenamefont
  {Zhang}}]{Zhao2016}%
  \BibitemOpen
  \bibfield  {author} {\bibinfo {author} {\bibfnamefont {X.}~\bibnamefont
  {Zhao}}, \bibinfo {author} {\bibfnamefont {C.}~\bibnamefont {Jin}}, \bibinfo
  {author} {\bibfnamefont {C.}~\bibnamefont {Wang}}, \bibinfo {author}
  {\bibfnamefont {H.}~\bibnamefont {Du}}, \bibinfo {author} {\bibfnamefont
  {J.}~\bibnamefont {Zang}}, \bibinfo {author} {\bibfnamefont {M.}~\bibnamefont
  {Tian}}, \bibinfo {author} {\bibfnamefont {R.}~\bibnamefont {Che}}, \ and\
  \bibinfo {author} {\bibfnamefont {Y.}~\bibnamefont {Zhang}},\ }\href
  {\doibase 10.1073/pnas.1600197113} {\bibfield  {journal} {\bibinfo  {journal}
  {Proc. Natl. Acad. Sci. U. S. A.}\ }\textbf {\bibinfo {volume} {113}},\
  \bibinfo {pages} {4918} (\bibinfo {year} {2016})}\BibitemShut {NoStop}%
\bibitem [{\citenamefont {Pfleiderer}\ \emph {et~al.}(2004)\citenamefont
  {Pfleiderer}, \citenamefont {Reznik}, \citenamefont {Pintschovius},
  \citenamefont {L{\"{o}}hneysen}, \citenamefont {Garst},\ and\ \citenamefont
  {Rosch}}]{Pfleiderer2004}%
  \BibitemOpen
  \bibfield  {author} {\bibinfo {author} {\bibfnamefont {C.}~\bibnamefont
  {Pfleiderer}}, \bibinfo {author} {\bibfnamefont {D.}~\bibnamefont {Reznik}},
  \bibinfo {author} {\bibfnamefont {L.}~\bibnamefont {Pintschovius}}, \bibinfo
  {author} {\bibfnamefont {H.~v.}\ \bibnamefont {L{\"{o}}hneysen}}, \bibinfo
  {author} {\bibfnamefont {M.}~\bibnamefont {Garst}}, \ and\ \bibinfo {author}
  {\bibfnamefont {A.}~\bibnamefont {Rosch}},\ }\href {\doibase
  10.1038/nature02232} {\bibfield  {journal} {\bibinfo  {journal} {Nature}\
  }\textbf {\bibinfo {volume} {427}},\ \bibinfo {pages} {227} (\bibinfo {year}
  {2004})}\BibitemShut {NoStop}%
\bibitem [{\citenamefont {Dup{\'{e}}}\ \emph {et~al.}(2014)\citenamefont
  {Dup{\'{e}}}, \citenamefont {Hoffmann}, \citenamefont {Paillard},\ and\
  \citenamefont {Heinze}}]{Dupe2014}%
  \BibitemOpen
  \bibfield  {author} {\bibinfo {author} {\bibfnamefont {B.}~\bibnamefont
  {Dup{\'{e}}}}, \bibinfo {author} {\bibfnamefont {M.}~\bibnamefont
  {Hoffmann}}, \bibinfo {author} {\bibfnamefont {C.}~\bibnamefont {Paillard}},
  \ and\ \bibinfo {author} {\bibfnamefont {S.}~\bibnamefont {Heinze}},\ }\href
  {\doibase 10.1038/ncomms5030} {\bibfield  {journal} {\bibinfo  {journal}
  {Nat. Commun.}\ }\textbf {\bibinfo {volume} {5}},\ \bibinfo {pages} {4030}
  (\bibinfo {year} {2014})}\BibitemShut {NoStop}%
\bibitem [{\citenamefont {Pollard}\ \emph {et~al.}(2017)\citenamefont
  {Pollard}, \citenamefont {Garlow}, \citenamefont {Yu}, \citenamefont {Wang},
  \citenamefont {Zhu},\ and\ \citenamefont {Yang}}]{Pollard2017}%
  \BibitemOpen
  \bibfield  {author} {\bibinfo {author} {\bibfnamefont {S.~D.}\ \bibnamefont
  {Pollard}}, \bibinfo {author} {\bibfnamefont {J.~A.}\ \bibnamefont {Garlow}},
  \bibinfo {author} {\bibfnamefont {J.}~\bibnamefont {Yu}}, \bibinfo {author}
  {\bibfnamefont {Z.}~\bibnamefont {Wang}}, \bibinfo {author} {\bibfnamefont
  {Y.}~\bibnamefont {Zhu}}, \ and\ \bibinfo {author} {\bibfnamefont
  {H.}~\bibnamefont {Yang}},\ }\href {\doibase 10.1038/ncomms14761} {\bibfield
  {journal} {\bibinfo  {journal} {Nat. Commun.}\ }\textbf {\bibinfo {volume}
  {8}},\ \bibinfo {pages} {14761} (\bibinfo {year} {2017})}\BibitemShut
  {NoStop}%
\bibitem [{\citenamefont {Soumyanarayanan}\ \emph {et~al.}(2017)\citenamefont
  {Soumyanarayanan}, \citenamefont {Raju}, \citenamefont {{Gonzalez Oyarce}},
  \citenamefont {Tan}, \citenamefont {Im}, \citenamefont {Petrovi{\'{c}}},
  \citenamefont {Ho}, \citenamefont {Khoo}, \citenamefont {Tran}, \citenamefont
  {Gan}, \citenamefont {Ernult},\ and\ \citenamefont
  {Panagopoulos}}]{Soumyanarayanan2017}%
  \BibitemOpen
  \bibfield  {author} {\bibinfo {author} {\bibfnamefont {A.}~\bibnamefont
  {Soumyanarayanan}}, \bibinfo {author} {\bibfnamefont {M.}~\bibnamefont
  {Raju}}, \bibinfo {author} {\bibfnamefont {A.~L.}\ \bibnamefont {{Gonzalez
  Oyarce}}}, \bibinfo {author} {\bibfnamefont {A.~K.~C.}\ \bibnamefont {Tan}},
  \bibinfo {author} {\bibfnamefont {M.-Y.}\ \bibnamefont {Im}}, \bibinfo
  {author} {\bibfnamefont {A.}~\bibnamefont {Petrovi{\'{c}}}}, \bibinfo
  {author} {\bibfnamefont {P.}~\bibnamefont {Ho}}, \bibinfo {author}
  {\bibfnamefont {K.~H.}\ \bibnamefont {Khoo}}, \bibinfo {author}
  {\bibfnamefont {M.}~\bibnamefont {Tran}}, \bibinfo {author} {\bibfnamefont
  {C.~K.}\ \bibnamefont {Gan}}, \bibinfo {author} {\bibfnamefont
  {F.}~\bibnamefont {Ernult}}, \ and\ \bibinfo {author} {\bibfnamefont
  {C.}~\bibnamefont {Panagopoulos}},\ }\href {\doibase 10.1038/nmat4934}
  {\bibfield  {journal} {\bibinfo  {journal} {Nat. Mater.}\ }\textbf {\bibinfo
  {volume} {16}},\ \bibinfo {pages} {898} (\bibinfo {year} {2017})}\BibitemShut
  {NoStop}%
\bibitem [{\citenamefont {Meyer}\ \emph {et~al.}(2019)\citenamefont {Meyer},
  \citenamefont {Perini}, \citenamefont {von Malottki}, \citenamefont
  {Kubetzka}, \citenamefont {Wiesendanger}, \citenamefont {von Bergmann},\ and\
  \citenamefont {Heinze}}]{Meyer2019}%
  \BibitemOpen
  \bibfield  {author} {\bibinfo {author} {\bibfnamefont {S.}~\bibnamefont
  {Meyer}}, \bibinfo {author} {\bibfnamefont {M.}~\bibnamefont {Perini}},
  \bibinfo {author} {\bibfnamefont {S.}~\bibnamefont {von Malottki}}, \bibinfo
  {author} {\bibfnamefont {A.}~\bibnamefont {Kubetzka}}, \bibinfo {author}
  {\bibfnamefont {R.}~\bibnamefont {Wiesendanger}}, \bibinfo {author}
  {\bibfnamefont {K.}~\bibnamefont {von Bergmann}}, \ and\ \bibinfo {author}
  {\bibfnamefont {S.}~\bibnamefont {Heinze}},\ }\href {\doibase
  10.1038/s41467-019-11831-4} {\bibfield  {journal} {\bibinfo  {journal} {Nat.
  Commun.}\ }\textbf {\bibinfo {volume} {10}},\ \bibinfo {pages} {3823}
  (\bibinfo {year} {2019})}\BibitemShut {NoStop}%
\bibitem [{\citenamefont {Farrell}\ and\ \citenamefont
  {Pereg-Barnea}(2014)}]{Farrell2014a}%
  \BibitemOpen
  \bibfield  {author} {\bibinfo {author} {\bibfnamefont {A.}~\bibnamefont
  {Farrell}}\ and\ \bibinfo {author} {\bibfnamefont {T.}~\bibnamefont
  {Pereg-Barnea}},\ }\href {\doibase 10.1103/PhysRevB.89.035112} {\bibfield
  {journal} {\bibinfo  {journal} {Phys. Rev. B}\ }\textbf {\bibinfo {volume}
  {89}},\ \bibinfo {pages} {035112} (\bibinfo {year} {2014})}\BibitemShut
  {NoStop}%
\bibitem [{\citenamefont {Chen}\ \emph {et~al.}(2016)\citenamefont {Chen},
  \citenamefont {Zhang},\ and\ \citenamefont {Liu}}]{Chen2016}%
  \BibitemOpen
  \bibfield  {author} {\bibinfo {author} {\bibfnamefont {J.~P.}\ \bibnamefont
  {Chen}}, \bibinfo {author} {\bibfnamefont {D.-W.}\ \bibnamefont {Zhang}}, \
  and\ \bibinfo {author} {\bibfnamefont {J.~M.}\ \bibnamefont {Liu}},\ }\href
  {\doibase 10.1038/srep29126} {\bibfield  {journal} {\bibinfo  {journal} {Sci.
  Rep.}\ }\textbf {\bibinfo {volume} {6}},\ \bibinfo {pages} {29126} (\bibinfo
  {year} {2016})}\BibitemShut {NoStop}%
\bibitem [{\citenamefont {R{\"{o}}{\ss}ler}\ \emph {et~al.}(2010)\citenamefont
  {R{\"{o}}{\ss}ler}, \citenamefont {Leonov},\ and\ \citenamefont
  {Bogdanov}}]{Rossler2010}%
  \BibitemOpen
  \bibfield  {author} {\bibinfo {author} {\bibfnamefont {U.~K.}\ \bibnamefont
  {R{\"{o}}{\ss}ler}}, \bibinfo {author} {\bibfnamefont {A.~A.}\ \bibnamefont
  {Leonov}}, \ and\ \bibinfo {author} {\bibfnamefont {A.~N.}\ \bibnamefont
  {Bogdanov}},\ }\href {\doibase 10.1088/1742-6596/200/2/022029} {\bibfield
  {journal} {\bibinfo  {journal} {J. Phys. Conf. Ser.}\ }\textbf {\bibinfo
  {volume} {200}},\ \bibinfo {pages} {022029} (\bibinfo {year}
  {2010})}\BibitemShut {NoStop}%
\bibitem [{\citenamefont {Ruderman}\ and\ \citenamefont
  {Kittel}(1954)}]{Ruderman1954}%
  \BibitemOpen
  \bibfield  {author} {\bibinfo {author} {\bibfnamefont {M.~A.}\ \bibnamefont
  {Ruderman}}\ and\ \bibinfo {author} {\bibfnamefont {C.}~\bibnamefont
  {Kittel}},\ }\href {\doibase 10.1103/PhysRev.96.99} {\bibfield  {journal}
  {\bibinfo  {journal} {Phys. Rev.}\ }\textbf {\bibinfo {volume} {96}},\
  \bibinfo {pages} {99} (\bibinfo {year} {1954})}\BibitemShut {NoStop}%
\bibitem [{\citenamefont {Kasuya}(1956)}]{Kasuya1956a}%
  \BibitemOpen
  \bibfield  {author} {\bibinfo {author} {\bibfnamefont {T.}~\bibnamefont
  {Kasuya}},\ }\href {\doibase 10.1143/PTP.16.45} {\bibfield  {journal}
  {\bibinfo  {journal} {Prog. Theor. Phys.}\ }\textbf {\bibinfo {volume}
  {16}},\ \bibinfo {pages} {45} (\bibinfo {year} {1956})}\BibitemShut {NoStop}%
\bibitem [{\citenamefont {Yosida}(1957)}]{Yosida1957a}%
  \BibitemOpen
  \bibfield  {author} {\bibinfo {author} {\bibfnamefont {K.}~\bibnamefont
  {Yosida}},\ }\href {\doibase 10.1103/PhysRev.106.893} {\bibfield  {journal}
  {\bibinfo  {journal} {Phys. Rev.}\ }\textbf {\bibinfo {volume} {106}},\
  \bibinfo {pages} {893} (\bibinfo {year} {1957})}\BibitemShut {NoStop}%
\bibitem [{\citenamefont {Hayami}\ and\ \citenamefont
  {Motome}(2018)}]{Hayami2018}%
  \BibitemOpen
  \bibfield  {author} {\bibinfo {author} {\bibfnamefont {S.}~\bibnamefont
  {Hayami}}\ and\ \bibinfo {author} {\bibfnamefont {Y.}~\bibnamefont
  {Motome}},\ }\href {\doibase 10.1103/PhysRevLett.121.137202} {\bibfield
  {journal} {\bibinfo  {journal} {Phys. Rev. Lett.}\ }\textbf {\bibinfo
  {volume} {121}},\ \bibinfo {pages} {137202} (\bibinfo {year}
  {2018})}\BibitemShut {NoStop}%
\bibitem [{\citenamefont {Bezvershenko}\ \emph {et~al.}(2018)\citenamefont
  {Bezvershenko}, \citenamefont {Kolezhuk},\ and\ \citenamefont
  {Ivanov}}]{Bezvershenko2018}%
  \BibitemOpen
  \bibfield  {author} {\bibinfo {author} {\bibfnamefont {A.~V.}\ \bibnamefont
  {Bezvershenko}}, \bibinfo {author} {\bibfnamefont {A.~K.}\ \bibnamefont
  {Kolezhuk}}, \ and\ \bibinfo {author} {\bibfnamefont {B.~A.}\ \bibnamefont
  {Ivanov}},\ }\href {\doibase 10.1103/PhysRevB.97.054408} {\bibfield
  {journal} {\bibinfo  {journal} {Phys. Rev. B}\ }\textbf {\bibinfo {volume}
  {97}},\ \bibinfo {pages} {054408} (\bibinfo {year} {2018})}\BibitemShut
  {NoStop}%
\bibitem [{\citenamefont {Seki}\ \emph
  {et~al.}(2012{\natexlab{a}})\citenamefont {Seki}, \citenamefont {Kim},
  \citenamefont {Inosov}, \citenamefont {Georgii}, \citenamefont {Keimer},
  \citenamefont {Ishiwata},\ and\ \citenamefont {Tokura}}]{Seki2012b}%
  \BibitemOpen
  \bibfield  {author} {\bibinfo {author} {\bibfnamefont {S.}~\bibnamefont
  {Seki}}, \bibinfo {author} {\bibfnamefont {J.-H.}\ \bibnamefont {Kim}},
  \bibinfo {author} {\bibfnamefont {D.~S.}\ \bibnamefont {Inosov}}, \bibinfo
  {author} {\bibfnamefont {R.}~\bibnamefont {Georgii}}, \bibinfo {author}
  {\bibfnamefont {B.}~\bibnamefont {Keimer}}, \bibinfo {author} {\bibfnamefont
  {S.}~\bibnamefont {Ishiwata}}, \ and\ \bibinfo {author} {\bibfnamefont
  {Y.}~\bibnamefont {Tokura}},\ }\href {\doibase 10.1103/PhysRevB.85.220406}
  {\bibfield  {journal} {\bibinfo  {journal} {Phys. Rev. B}\ }\textbf {\bibinfo
  {volume} {85}},\ \bibinfo {pages} {220406} (\bibinfo {year}
  {2012}{\natexlab{a}})}\BibitemShut {NoStop}%
\bibitem [{\citenamefont {Seki}\ \emph
  {et~al.}(2012{\natexlab{b}})\citenamefont {Seki}, \citenamefont {Yu},
  \citenamefont {Ishiwata},\ and\ \citenamefont {Tokura}}]{Seki2012c}%
  \BibitemOpen
  \bibfield  {author} {\bibinfo {author} {\bibfnamefont {S.}~\bibnamefont
  {Seki}}, \bibinfo {author} {\bibfnamefont {X.~Z.}\ \bibnamefont {Yu}},
  \bibinfo {author} {\bibfnamefont {S.}~\bibnamefont {Ishiwata}}, \ and\
  \bibinfo {author} {\bibfnamefont {Y.}~\bibnamefont {Tokura}},\ }\href
  {\doibase 10.1126/science.1214143} {\bibfield  {journal} {\bibinfo  {journal}
  {Science}\ }\textbf {\bibinfo {volume} {336}},\ \bibinfo {pages} {198}
  (\bibinfo {year} {2012}{\natexlab{b}})}\BibitemShut {NoStop}%
\bibitem [{\citenamefont {Dzyaloshinsky}(1958)}]{Dzyaloshinsky1958}%
  \BibitemOpen
  \bibfield  {author} {\bibinfo {author} {\bibfnamefont {I.}~\bibnamefont
  {Dzyaloshinsky}},\ }\href {\doibase 10.1016/0022-3697(58)90076-3} {\bibfield
  {journal} {\bibinfo  {journal} {J. Phys. Chem. Solids}\ }\textbf {\bibinfo
  {volume} {4}},\ \bibinfo {pages} {241} (\bibinfo {year} {1958})}\BibitemShut
  {NoStop}%
\bibitem [{\citenamefont {Moriya}(1960)}]{Moriya1960}%
  \BibitemOpen
  \bibfield  {author} {\bibinfo {author} {\bibfnamefont {T.}~\bibnamefont
  {Moriya}},\ }\href {\doibase 10.1103/PhysRev.120.91} {\bibfield  {journal}
  {\bibinfo  {journal} {Phys. Rev.}\ }\textbf {\bibinfo {volume} {120}},\
  \bibinfo {pages} {91} (\bibinfo {year} {1960})}\BibitemShut {NoStop}%
\bibitem [{\citenamefont {R{\"{o}}{\ss}ler}\ \emph {et~al.}(2006)\citenamefont
  {R{\"{o}}{\ss}ler}, \citenamefont {Bogdanov},\ and\ \citenamefont
  {Pfleiderer}}]{Rossler2006}%
  \BibitemOpen
  \bibfield  {author} {\bibinfo {author} {\bibfnamefont {U.~K.}\ \bibnamefont
  {R{\"{o}}{\ss}ler}}, \bibinfo {author} {\bibfnamefont {A.~N.}\ \bibnamefont
  {Bogdanov}}, \ and\ \bibinfo {author} {\bibfnamefont {C.}~\bibnamefont
  {Pfleiderer}},\ }\href {\doibase 10.1038/nature05056} {\bibfield  {journal}
  {\bibinfo  {journal} {Nature}\ }\textbf {\bibinfo {volume} {442}},\ \bibinfo
  {pages} {797} (\bibinfo {year} {2006})}\BibitemShut {NoStop}%
\bibitem [{\citenamefont {Dagotto}(2002)}]{Dagotto2002}%
  \BibitemOpen
  \bibfield  {author} {\bibinfo {author} {\bibfnamefont {E.}~\bibnamefont
  {Dagotto}},\ }\href@noop {} {\emph {\bibinfo {title} {{Nanoscale Phase
  Separation and Collosal Magnetoresistance}}}}\ (\bibinfo  {publisher}
  {Springer Berlin/Heidelberg},\ \bibinfo {address} {Berlin},\ \bibinfo {year}
  {2002})\BibitemShut {NoStop}%
\bibitem [{\citenamefont {Alvarez}\ \emph {et~al.}(2002)\citenamefont
  {Alvarez}, \citenamefont {Mayr},\ and\ \citenamefont
  {Dagotto}}]{Alvarez2002}%
  \BibitemOpen
  \bibfield  {author} {\bibinfo {author} {\bibfnamefont {G.}~\bibnamefont
  {Alvarez}}, \bibinfo {author} {\bibfnamefont {M.}~\bibnamefont {Mayr}}, \
  and\ \bibinfo {author} {\bibfnamefont {E.}~\bibnamefont {Dagotto}},\ }\href
  {\doibase 10.1103/PhysRevLett.89.277202} {\bibfield  {journal} {\bibinfo
  {journal} {Phys. Rev. Lett.}\ }\textbf {\bibinfo {volume} {89}},\ \bibinfo
  {pages} {277202} (\bibinfo {year} {2002})}\BibitemShut {NoStop}%
\bibitem [{\citenamefont {Berciu}\ and\ \citenamefont
  {Bhatt}(2001)}]{Berciu2001}%
  \BibitemOpen
  \bibfield  {author} {\bibinfo {author} {\bibfnamefont {M.}~\bibnamefont
  {Berciu}}\ and\ \bibinfo {author} {\bibfnamefont {R.~N.}\ \bibnamefont
  {Bhatt}},\ }\href {\doibase 10.1103/PhysRevLett.87.107203} {\bibfield
  {journal} {\bibinfo  {journal} {Phys. Rev. Lett.}\ }\textbf {\bibinfo
  {volume} {87}},\ \bibinfo {pages} {107203} (\bibinfo {year}
  {2001})}\BibitemShut {NoStop}%
\bibitem [{\citenamefont {Pradhan}\ and\ \citenamefont
  {Das}(2017)}]{Pradhan2017}%
  \BibitemOpen
  \bibfield  {author} {\bibinfo {author} {\bibfnamefont {K.}~\bibnamefont
  {Pradhan}}\ and\ \bibinfo {author} {\bibfnamefont {S.~K.}\ \bibnamefont
  {Das}},\ }\href {\doibase 10.1038/s41598-017-09729-6} {\bibfield  {journal}
  {\bibinfo  {journal} {Sci. Rep.}\ }\textbf {\bibinfo {volume} {7}},\ \bibinfo
  {pages} {9603} (\bibinfo {year} {2017})}\BibitemShut {NoStop}%
\bibitem [{\citenamefont {Yaouanc}\ \emph {et~al.}(2020)\citenamefont
  {Yaouanc}, \citenamefont {{Dalmas de R{\'{e}}otier}}, \citenamefont
  {Roessli}, \citenamefont {Maisuradze}, \citenamefont {Amato}, \citenamefont
  {Andreica},\ and\ \citenamefont {Lapertot}}]{Yaouanc2020}%
  \BibitemOpen
  \bibfield  {author} {\bibinfo {author} {\bibfnamefont {A.}~\bibnamefont
  {Yaouanc}}, \bibinfo {author} {\bibfnamefont {P.}~\bibnamefont {{Dalmas de
  R{\'{e}}otier}}}, \bibinfo {author} {\bibfnamefont {B.}~\bibnamefont
  {Roessli}}, \bibinfo {author} {\bibfnamefont {A.}~\bibnamefont {Maisuradze}},
  \bibinfo {author} {\bibfnamefont {A.}~\bibnamefont {Amato}}, \bibinfo
  {author} {\bibfnamefont {D.}~\bibnamefont {Andreica}}, \ and\ \bibinfo
  {author} {\bibfnamefont {G.}~\bibnamefont {Lapertot}},\ }\href {\doibase
  10.1103/PhysRevResearch.2.013029} {\bibfield  {journal} {\bibinfo  {journal}
  {Phys. Rev. Res.}\ }\textbf {\bibinfo {volume} {2}},\ \bibinfo {pages}
  {013029} (\bibinfo {year} {2020})}\BibitemShut {NoStop}%
\bibitem [{\citenamefont {Bombor}\ \emph {et~al.}(2013)\citenamefont {Bombor},
  \citenamefont {Blum}, \citenamefont {Volkonskiy}, \citenamefont {Rodan},
  \citenamefont {Wurmehl}, \citenamefont {Hess},\ and\ \citenamefont
  {B{\"{u}}chner}}]{Bombor2013}%
  \BibitemOpen
  \bibfield  {author} {\bibinfo {author} {\bibfnamefont {D.}~\bibnamefont
  {Bombor}}, \bibinfo {author} {\bibfnamefont {C.~G.~F.}\ \bibnamefont {Blum}},
  \bibinfo {author} {\bibfnamefont {O.}~\bibnamefont {Volkonskiy}}, \bibinfo
  {author} {\bibfnamefont {S.}~\bibnamefont {Rodan}}, \bibinfo {author}
  {\bibfnamefont {S.}~\bibnamefont {Wurmehl}}, \bibinfo {author} {\bibfnamefont
  {C.}~\bibnamefont {Hess}}, \ and\ \bibinfo {author} {\bibfnamefont
  {B.}~\bibnamefont {B{\"{u}}chner}},\ }\href {\doibase
  10.1103/PhysRevLett.110.066601} {\bibfield  {journal} {\bibinfo  {journal}
  {Phys. Rev. Lett.}\ }\textbf {\bibinfo {volume} {110}},\ \bibinfo {pages}
  {066601} (\bibinfo {year} {2013})}\BibitemShut {NoStop}%
\bibitem [{\citenamefont {Felser}\ \emph {et~al.}(2015)\citenamefont {Felser},
  \citenamefont {Wollmann}, \citenamefont {Chadov}, \citenamefont {Fecher},\
  and\ \citenamefont {Parkin}}]{Felser2015}%
  \BibitemOpen
  \bibfield  {author} {\bibinfo {author} {\bibfnamefont {C.}~\bibnamefont
  {Felser}}, \bibinfo {author} {\bibfnamefont {L.}~\bibnamefont {Wollmann}},
  \bibinfo {author} {\bibfnamefont {S.}~\bibnamefont {Chadov}}, \bibinfo
  {author} {\bibfnamefont {G.~H.}\ \bibnamefont {Fecher}}, \ and\ \bibinfo
  {author} {\bibfnamefont {S.~S.~P.}\ \bibnamefont {Parkin}},\ }\href {\doibase
  10.1063/1.4917387} {\bibfield  {journal} {\bibinfo  {journal} {APL Mater.}\
  }\textbf {\bibinfo {volume} {3}},\ \bibinfo {pages} {041518} (\bibinfo {year}
  {2015})}\BibitemShut {NoStop}%
\bibitem [{\citenamefont {Şaşıoğlu}\ \emph {et~al.}(2008)\citenamefont
  {Şaşıoğlu}, \citenamefont {Sandratskii},\ and\ \citenamefont
  {Bruno}}]{Sasoglu2008}%
  \BibitemOpen
  \bibfield  {author} {\bibinfo {author} {\bibfnamefont {E.}~\bibnamefont
  {Şaşıoğlu}}, \bibinfo {author} {\bibfnamefont {L.~M.}\ \bibnamefont
  {Sandratskii}}, \ and\ \bibinfo {author} {\bibfnamefont {P.}~\bibnamefont
  {Bruno}},\ }\href {\doibase 10.1103/PhysRevB.77.064417} {\bibfield  {journal}
  {\bibinfo  {journal} {Phys. Rev. B}\ }\textbf {\bibinfo {volume} {77}},\
  \bibinfo {pages} {064417} (\bibinfo {year} {2008})}\BibitemShut {NoStop}%
\bibitem [{\citenamefont {Okada}\ \emph {et~al.}(2018)\citenamefont {Okada},
  \citenamefont {Kato},\ and\ \citenamefont {Motome}}]{Okada2018}%
  \BibitemOpen
  \bibfield  {author} {\bibinfo {author} {\bibfnamefont {K.~N.}\ \bibnamefont
  {Okada}}, \bibinfo {author} {\bibfnamefont {Y.}~\bibnamefont {Kato}}, \ and\
  \bibinfo {author} {\bibfnamefont {Y.}~\bibnamefont {Motome}},\ }\href
  {\doibase 10.1103/PhysRevB.98.224406} {\bibfield  {journal} {\bibinfo
  {journal} {Phys. Rev. B}\ }\textbf {\bibinfo {volume} {98}},\ \bibinfo
  {pages} {224406} (\bibinfo {year} {2018})}\BibitemShut {NoStop}%
\bibitem [{SM()}]{SM}%
  \BibitemOpen
  \href@noop {} {\enquote {\bibinfo {title} {See the appended supplemental material}}\ }\BibitemShut {NoStop}%
\bibitem [{\citenamefont {Yunoki}\ \emph {et~al.}(1998)\citenamefont {Yunoki},
  \citenamefont {Hu}, \citenamefont {Malvezzi}, \citenamefont {Moreo},
  \citenamefont {Furukawa},\ and\ \citenamefont {Dagotto}}]{Yunoki1998b}%
  \BibitemOpen
  \bibfield  {author} {\bibinfo {author} {\bibfnamefont {S.}~\bibnamefont
  {Yunoki}}, \bibinfo {author} {\bibfnamefont {J.}~\bibnamefont {Hu}}, \bibinfo
  {author} {\bibfnamefont {A.~L.}\ \bibnamefont {Malvezzi}}, \bibinfo {author}
  {\bibfnamefont {A.}~\bibnamefont {Moreo}}, \bibinfo {author} {\bibfnamefont
  {N.}~\bibnamefont {Furukawa}}, \ and\ \bibinfo {author} {\bibfnamefont
  {E.}~\bibnamefont {Dagotto}},\ }\href {\doibase 10.1103/PhysRevLett.80.845}
  {\bibfield  {journal} {\bibinfo  {journal} {Phys. Rev. Lett.}\ }\textbf
  {\bibinfo {volume} {80}},\ \bibinfo {pages} {845} (\bibinfo {year}
  {1998})}\BibitemShut {NoStop}%
\bibitem [{\citenamefont {Kumar}\ and\ \citenamefont
  {Majumdar}(2005)}]{Kumar2005a}%
  \BibitemOpen
  \bibfield  {author} {\bibinfo {author} {\bibfnamefont {S.}~\bibnamefont
  {Kumar}}\ and\ \bibinfo {author} {\bibfnamefont {P.}~\bibnamefont
  {Majumdar}},\ }\href {\doibase 10.1140/epjb/e2005-00261-9} {\bibfield
  {journal} {\bibinfo  {journal} {Eur. Phys. J. B}\ }\textbf {\bibinfo {volume}
  {46}},\ \bibinfo {pages} {315} (\bibinfo {year} {2005})}\BibitemShut
  {NoStop}%
\bibitem [{\citenamefont {Calder{\'{o}}n}\ and\ \citenamefont
  {Brey}(1998)}]{Calderon1998a}%
  \BibitemOpen
  \bibfield  {author} {\bibinfo {author} {\bibfnamefont {M.~J.}\ \bibnamefont
  {Calder{\'{o}}n}}\ and\ \bibinfo {author} {\bibfnamefont {L.}~\bibnamefont
  {Brey}},\ }\href {\doibase 10.1103/PhysRevB.58.3286} {\bibfield  {journal}
  {\bibinfo  {journal} {Phys. Rev. B}\ }\textbf {\bibinfo {volume} {58}},\
  \bibinfo {pages} {3286} (\bibinfo {year} {1998})}\BibitemShut {NoStop}%
\bibitem [{\citenamefont {Tokiwa}\ \emph {et~al.}(2014)\citenamefont {Tokiwa},
  \citenamefont {Ishikawa}, \citenamefont {Nakatsuji},\ and\ \citenamefont
  {Gegenwart}}]{Tokiwa2014}%
  \BibitemOpen
  \bibfield  {author} {\bibinfo {author} {\bibfnamefont {Y.}~\bibnamefont
  {Tokiwa}}, \bibinfo {author} {\bibfnamefont {J.~J.}\ \bibnamefont
  {Ishikawa}}, \bibinfo {author} {\bibfnamefont {S.}~\bibnamefont {Nakatsuji}},
  \ and\ \bibinfo {author} {\bibfnamefont {P.}~\bibnamefont {Gegenwart}},\
  }\href {\doibase 10.1038/nmat3900} {\bibfield  {journal} {\bibinfo  {journal}
  {Nat. Mater.}\ }\textbf {\bibinfo {volume} {13}},\ \bibinfo {pages} {356}
  (\bibinfo {year} {2014})}\BibitemShut {NoStop}%
\bibitem [{\citenamefont {Okabe}\ \emph {et~al.}(2019)\citenamefont {Okabe},
  \citenamefont {Hiraishi}, \citenamefont {Koda}, \citenamefont {Kojima},
  \citenamefont {Takeshita}, \citenamefont {Yamauchi}, \citenamefont
  {Matsushita}, \citenamefont {Kuramoto},\ and\ \citenamefont
  {Kadono}}]{Okabe2019}%
  \BibitemOpen
  \bibfield  {author} {\bibinfo {author} {\bibfnamefont {H.}~\bibnamefont
  {Okabe}}, \bibinfo {author} {\bibfnamefont {M.}~\bibnamefont {Hiraishi}},
  \bibinfo {author} {\bibfnamefont {A.}~\bibnamefont {Koda}}, \bibinfo {author}
  {\bibfnamefont {K.~M.}\ \bibnamefont {Kojima}}, \bibinfo {author}
  {\bibfnamefont {S.}~\bibnamefont {Takeshita}}, \bibinfo {author}
  {\bibfnamefont {I.}~\bibnamefont {Yamauchi}}, \bibinfo {author}
  {\bibfnamefont {Y.}~\bibnamefont {Matsushita}}, \bibinfo {author}
  {\bibfnamefont {Y.}~\bibnamefont {Kuramoto}}, \ and\ \bibinfo {author}
  {\bibfnamefont {R.}~\bibnamefont {Kadono}},\ }\href {\doibase
  10.1103/PhysRevB.99.041113} {\bibfield  {journal} {\bibinfo  {journal} {Phys.
  Rev. B}\ }\textbf {\bibinfo {volume} {99}},\ \bibinfo {pages} {041113}
  (\bibinfo {year} {2019})}\BibitemShut {NoStop}%
\bibitem [{\citenamefont {Nakatsuji}\ \emph {et~al.}(2006)\citenamefont
  {Nakatsuji}, \citenamefont {Machida}, \citenamefont {Maeno}, \citenamefont
  {Tayama}, \citenamefont {Sakakibara}, \citenamefont {van Duijn},
  \citenamefont {Balicas}, \citenamefont {Millican}, \citenamefont {Macaluso},\
  and\ \citenamefont {Chan}}]{Nakatsuji2006}%
  \BibitemOpen
  \bibfield  {author} {\bibinfo {author} {\bibfnamefont {S.}~\bibnamefont
  {Nakatsuji}}, \bibinfo {author} {\bibfnamefont {Y.}~\bibnamefont {Machida}},
  \bibinfo {author} {\bibfnamefont {Y.}~\bibnamefont {Maeno}}, \bibinfo
  {author} {\bibfnamefont {T.}~\bibnamefont {Tayama}}, \bibinfo {author}
  {\bibfnamefont {T.}~\bibnamefont {Sakakibara}}, \bibinfo {author}
  {\bibfnamefont {J.}~\bibnamefont {van Duijn}}, \bibinfo {author}
  {\bibfnamefont {L.}~\bibnamefont {Balicas}}, \bibinfo {author} {\bibfnamefont
  {J.~N.}\ \bibnamefont {Millican}}, \bibinfo {author} {\bibfnamefont {R.~T.}\
  \bibnamefont {Macaluso}}, \ and\ \bibinfo {author} {\bibfnamefont {J.~Y.}\
  \bibnamefont {Chan}},\ }\href {\doibase 10.1103/PhysRevLett.96.087204}
  {\bibfield  {journal} {\bibinfo  {journal} {Phys. Rev. Lett.}\ }\textbf
  {\bibinfo {volume} {96}},\ \bibinfo {pages} {087204} (\bibinfo {year}
  {2006})}\BibitemShut {NoStop}%
\bibitem [{\citenamefont {McKeever}\ \emph {et~al.}(2019)\citenamefont
  {McKeever}, \citenamefont {Rodrigues}, \citenamefont {Pinna}, \citenamefont
  {Abanov}, \citenamefont {Sinova},\ and\ \citenamefont
  {Everschor-Sitte}}]{McKeever2019}%
  \BibitemOpen
  \bibfield  {author} {\bibinfo {author} {\bibfnamefont {B.~F.}\ \bibnamefont
  {McKeever}}, \bibinfo {author} {\bibfnamefont {D.~R.}\ \bibnamefont
  {Rodrigues}}, \bibinfo {author} {\bibfnamefont {D.}~\bibnamefont {Pinna}},
  \bibinfo {author} {\bibfnamefont {A.}~\bibnamefont {Abanov}}, \bibinfo
  {author} {\bibfnamefont {J.}~\bibnamefont {Sinova}}, \ and\ \bibinfo {author}
  {\bibfnamefont {K.}~\bibnamefont {Everschor-Sitte}},\ }\href {\doibase
  10.1103/PhysRevB.99.054430} {\bibfield  {journal} {\bibinfo  {journal} {Phys.
  Rev. B}\ }\textbf {\bibinfo {volume} {99}},\ \bibinfo {pages} {054430}
  (\bibinfo {year} {2019})}\BibitemShut {NoStop}%
\bibitem [{\citenamefont {Woo}\ \emph {et~al.}(2017)\citenamefont {Woo},
  \citenamefont {Song}, \citenamefont {Han}, \citenamefont {Jung},
  \citenamefont {Im}, \citenamefont {Lee}, \citenamefont {Song}, \citenamefont
  {Fischer}, \citenamefont {Hong}, \citenamefont {Choi}, \citenamefont {Min},
  \citenamefont {Koo},\ and\ \citenamefont {Chang}}]{Woo2017}%
  \BibitemOpen
  \bibfield  {author} {\bibinfo {author} {\bibfnamefont {S.}~\bibnamefont
  {Woo}}, \bibinfo {author} {\bibfnamefont {K.~M.}\ \bibnamefont {Song}},
  \bibinfo {author} {\bibfnamefont {H.-S.}\ \bibnamefont {Han}}, \bibinfo
  {author} {\bibfnamefont {M.-S.}\ \bibnamefont {Jung}}, \bibinfo {author}
  {\bibfnamefont {M.-Y.}\ \bibnamefont {Im}}, \bibinfo {author} {\bibfnamefont
  {K.-S.}\ \bibnamefont {Lee}}, \bibinfo {author} {\bibfnamefont {K.~S.}\
  \bibnamefont {Song}}, \bibinfo {author} {\bibfnamefont {P.}~\bibnamefont
  {Fischer}}, \bibinfo {author} {\bibfnamefont {J.-I.}\ \bibnamefont {Hong}},
  \bibinfo {author} {\bibfnamefont {J.~W.}\ \bibnamefont {Choi}}, \bibinfo
  {author} {\bibfnamefont {B.-C.}\ \bibnamefont {Min}}, \bibinfo {author}
  {\bibfnamefont {H.~C.}\ \bibnamefont {Koo}}, \ and\ \bibinfo {author}
  {\bibfnamefont {J.}~\bibnamefont {Chang}},\ }\href {\doibase
  10.1038/ncomms15573} {\bibfield  {journal} {\bibinfo  {journal} {Nat.
  Commun.}\ }\textbf {\bibinfo {volume} {8}},\ \bibinfo {pages} {15573}
  (\bibinfo {year} {2017})}\BibitemShut {NoStop}%
\bibitem [{\citenamefont {Martin}\ and\ \citenamefont
  {Batista}(2008)}]{Martin2008a}%
  \BibitemOpen
  \bibfield  {author} {\bibinfo {author} {\bibfnamefont {I.}~\bibnamefont
  {Martin}}\ and\ \bibinfo {author} {\bibfnamefont {C.~D.}\ \bibnamefont
  {Batista}},\ }\href {\doibase 10.1103/PhysRevLett.101.156402} {\bibfield
  {journal} {\bibinfo  {journal} {Phys. Rev. Lett.}\ }\textbf {\bibinfo
  {volume} {101}},\ \bibinfo {pages} {156402} (\bibinfo {year}
  {2008})}\BibitemShut {NoStop}%
\bibitem [{\citenamefont {Pasrija}\ and\ \citenamefont
  {Kumar}(2016)}]{Pasrija2016}%
  \BibitemOpen
  \bibfield  {author} {\bibinfo {author} {\bibfnamefont {K.}~\bibnamefont
  {Pasrija}}\ and\ \bibinfo {author} {\bibfnamefont {S.}~\bibnamefont
  {Kumar}},\ }\href {\doibase 10.1103/PhysRevB.93.195110} {\bibfield  {journal}
  {\bibinfo  {journal} {Phys. Rev. B}\ }\textbf {\bibinfo {volume} {93}},\
  \bibinfo {pages} {195110} (\bibinfo {year} {2016})}\BibitemShut {NoStop}%
\end{thebibliography}
%

\newpage

\onecolumngrid
\begin{center}
{\bf {\Large{Supplemental Material}}}
\end{center}
\section{Derivation of the new spin Hamiltonian}
The ferromagnetic Kondo lattice model (FKLM) in the presence of Rashba coupling on a square lattice is described by the Hamiltonian,
\begin{eqnarray}
H & = & - t \sum_{\langle ij \rangle,\sigma} (c^\dagger_{i\sigma} c^{}_{j\sigma} + {\textrm H.c.}) 
+ \lambda \sum_{i} [(c^{\dagger}_{i \downarrow} c^{}_{i+x\uparrow} - c^{\dagger}_{i\uparrow} c^{}_{i+x\downarrow}) \nonumber \\
& & + \textrm{i} (c^{\dagger}_{i\downarrow} c^{}_{i+y\uparrow} + c^{\dagger}_{i\uparrow} c^{}_{i+y\downarrow}) + {\textrm H.c.}] - J_H \sum_{i} {\bf S}_i \cdot {\bf s}_i.
\label{eq:Ham}
\end{eqnarray}
\noindent
The notations remain identical to that used in the main text. In order to handle the large $J_H$ limit, we perform a site dependent rotation of the spin-$\frac{1}{2}$ basis given by the canonical $SU(2)$ transformation,

$\begin{bmatrix}
c_{i\uparrow} \\
c_{i\downarrow}
\end{bmatrix} 
=
\begin{bmatrix}
\cos(\frac {\theta_i}{2})   & -\sin(\frac {\theta_i}{2}) e^{-\textrm{i} \phi_i} \\
\sin(\frac {\theta_i}{2}) e^{\textrm{i} \phi_i} & \cos(\frac {\theta_i}{2}) 

\end{bmatrix}  \begin{bmatrix}
d_{ip} \\
d_{ia}
\end{bmatrix}$. \\
\\
Here, $d_{ip} (d_{ia})$  annihilates an electron at site ${i}$ with spin parallel (antiparallel) to the localized spin and $\theta_i$, $\phi_i$ are the polar and azimuthal angles describing the direction of the local spin ${\bf S}_i$. In the large $J_H$ limit, the low energy physics is retained in parallel subspace, leading to the Rashba double-exchange (RDE) Hamiltonian,
\\
\begin{eqnarray}
H_{{\rm RDE}} & = &  \sum_{\langle ij \rangle,\gamma} [g^{\gamma}_{ij} d^\dagger_{ip} d^{}_{jp} + {\textrm H.c.}],
\label{eq:Ham2}
\end{eqnarray}
\\
\noindent
where, site $j = i + \gamma$ is the nn of site $i$ along spatial direction $\gamma \in \{x,y\}$. The projected hopping parameters, $g^{\gamma}_{ij}= t^{\gamma}_{ij} + \lambda^{\gamma}_{ij}$ , are given by,

\begin{eqnarray}
t^{\gamma}_{ij} & = & -t \big[\cos(\frac {\theta_i}{2}) \cos(\frac {\theta_j}{2}) 
+ \sin(\frac {\theta_i}{2})  \sin(\frac {\theta_j}{2})e^{-\textrm{i} (\phi_i-\phi_j)}  \big] ,
\nonumber \\
\lambda_{{ij}}^x & = & \lambda \big[\sin(\frac {\theta_i}{2})  \cos(\frac {\theta_j}{2})e^{-\textrm{i} \phi_i} - \cos(\frac {\theta_i}{2})  \sin(\frac {\theta_j}{2})e^{\textrm{i} \phi_j}\big] ,
\nonumber \\
\lambda_{{ij}}^y & = & \textrm{i} \lambda \big[\sin(\frac {\theta_i}{2})  \cos(\frac {\theta_j}{2})e^{-\textrm{i} \phi_i} + \cos(\frac {\theta_i}{2})  \sin(\frac {\theta_j}{2})e^{\textrm{i} \phi_j}\big] .
\end{eqnarray}

\pagebreak

\noindent
Writing $g^{\gamma}_{ij}$ in polar form, $g^{\gamma}_{ij}= f^{\gamma}_{ij} e^{\textrm{i} h^{\gamma}_{ij}}$, we obtain the following closed form expressions for $f^{x}_{ij}$ and $f^{y}_{ij}$ :

\begin{eqnarray*}
f^{x}_{ij} & = &  \sqrt{\frac {1}{2}[t^2(1+S_i^{x}S_j^{x}+S_i^{y}S_j^{y}+S_i^{z}S_j^{z})+\lambda^2(1-S_i^{x}S_j^{x}+S_i^{y}S_j^{y}-S_i^{z}S_j^{z})-2t\lambda(S_i^{x}S_j^{z}-S_i^{z}S_j^{x})]} \nonumber \\
& = &  \sqrt{\frac {1}{2}[t^2(1+{\bf S}_i \cdot {\bf S}_j)+\lambda^2(1-{\bf S}_i \cdot {\bf S}_j+2S_i^{y}S_j^{y})+2t\lambda \hat{\bf y} \cdot ({\bf S}_i \times{\bf S}_j)]} ,
\end{eqnarray*}

\begin{eqnarray}   \label{eq:ESH} 
f^{y}_{ij} & = &  \sqrt{\frac {1}{2}[t^2(1+S_i^{x}S_j^{x}+S_i^{y}S_j^{y}+S_i^{z}S_j^{z})+\lambda^2(1+S_i^{x}S_j^{x}-S_i^{y}S_j^{y}-S_i^{z}S_j^{z})+2t\lambda(S_i^{z}S_j^{y}-S_i^{y}S_j^{z})]} \nonumber \\
& = &  \sqrt{\frac {1}{2}[t^2(1+{\bf S}_i \cdot {\bf S}_j)+\lambda^2(1-{\bf S}_i \cdot {\bf S}_j+2S_i^{x}S_j^{x})-2t\lambda \hat{\bf x} \cdot  ({\bf S}_i \times{\bf S}_j)]} .
\end{eqnarray} 
\\

\noindent
The phase angles, $h^{\gamma}_{ij}$, are easily obtained via,
\begin{eqnarray}
h^{\gamma}_{ij} & = &\arctan \left(\frac{\text{Im}(g^{\gamma}_{ij})}{\text{Re}(g^{\gamma}_{ij})}\right),
\end{eqnarray}
 
\noindent
where, real and imaginary parts of $g^{\gamma}_{ij}$ are given by,

\begin{align}
\text{Re}(g^{x}_{ij}) &=
\begin{aligned}[t]
&-t(\cos(\frac {\theta_i}{2}) \cos(\frac {\theta_j}{2})  + \sin(\frac {\theta_i}{2})  \sin(\frac {\theta_j}{2})\cos(\phi_i-\phi_j)) \\
&+\lambda(\sin(\frac {\theta_i}{2})  \cos(\frac {\theta_j}{2}) \cos\phi_i- \cos(\frac {\theta_i}{2})  \sin(\frac {\theta_j}{2})\cos\phi_j ),
\end{aligned} \\
\nonumber
\text{Im}(g^{x}_{ij}) &=
\begin{aligned}[t]
&t(\sin(\frac {\theta_i}{2})  \sin(\frac {\theta_j}{2})\sin(\phi_i-\phi_j))-\lambda(\sin(\frac {\theta_i}{2})  \cos(\frac {\theta_j}{2}) \sin\phi_i + \cos(\frac {\theta_i}{2})  \sin(\frac {\theta_j}{2})\sin\phi_j ),
\end{aligned} \\
\nonumber
\text{Re}(g^{y}_{ij}) &=
\begin{aligned}[t]
& -t(\cos(\frac {\theta_i}{2}) \cos(\frac {\theta_j}{2})  + \sin(\frac {\theta_i}{2})  \sin(\frac {\theta_j}{2})\cos(\phi_i-\phi_j)) \\
& - \lambda(\cos(\frac {\theta_i}{2})  \sin(\frac {\theta_j}{2}) \sin\phi_j- \sin(\frac {\theta_i}{2})  \cos(\frac {\theta_j}{2})\sin\phi_i ),
\end{aligned} \\
\nonumber
\text{Im}(g^{y}_{ij}) &=
\begin{aligned}
&  t(\sin(\frac {\theta_i}{2})  \sin(\frac {\theta_j}{2})\sin(\phi_i-\phi_j))+\lambda(\sin(\frac {\theta_i}{2})  \cos(\frac {\theta_j}{2}) \cos\phi_i + \cos(\frac {\theta_i}{2})  \sin(\frac {\theta_j}{2})\cos\phi_j).
\end{aligned}
\end{align}

\pagebreak
\noindent
The ground state expectation values of the Hamiltonian Eq. (\ref{eq:Ham2}) is identical to the expression,  $-\sum_{\langle ij \rangle,\gamma}   D^{\gamma}_{ij} f^{\gamma}_{ij}$ , where  $  D^{\gamma}_{ij} = \langle [e^{\textrm{i} h^{\gamma}_{ij}} d^\dagger_{ip} d^{}_{jp} + {\textrm H.c.}]\rangle_{gs} $. Following the strategy used in double exchange models, we promote the above expression to a spin Hamiltonian,

\begin{eqnarray}
H_{{\rm S}} & = & -\sum_{\langle ij \rangle,\gamma}   D^{\gamma}_{ij} f^{\gamma}_{ij} .
\label{eq:Heff}
\end{eqnarray}

\noindent
We emphasize that, by construction, the magnetic ground states of $H_{{\rm S}}$ Eq. (\ref{eq:Heff})  and $H_{{\rm RDE}}$ Eq. (\ref{eq:Ham2}) are identical.
\section{Distribution of coupling constants}
 
The coupling constants of the   effective Hamiltonian  Eq. (\ref{eq:Heff}) are determined as the expectation values in the ground states obtained via EDMC on $H_{{\rm RDE}}$. We calculate the distributions of $D^{\gamma}_{ij}$ for the pairs of nearest neighbor sites in different ground states obtained via EDMC. The density of $D^{\gamma}_{ij}$ is defined as,

\begin{eqnarray}
\mathcal{N}(D)  =  1/N \sum_{\langle ij \rangle} \delta(D - D^{\gamma}_{ij}) \nonumber 
 \approx  1/N \sum_{\langle ij \rangle} \frac{\eta/\pi}{\eta^2 + (D - D^{\gamma}_{ij})^2}, 
\end{eqnarray}

where, $\eta$ is Lorentzian broadening parameters which is set to $0.001$ for calculations. 
\begin{figure}[H]
	\centering
	\includegraphics[width=.45 \columnwidth,angle=0,clip=true]{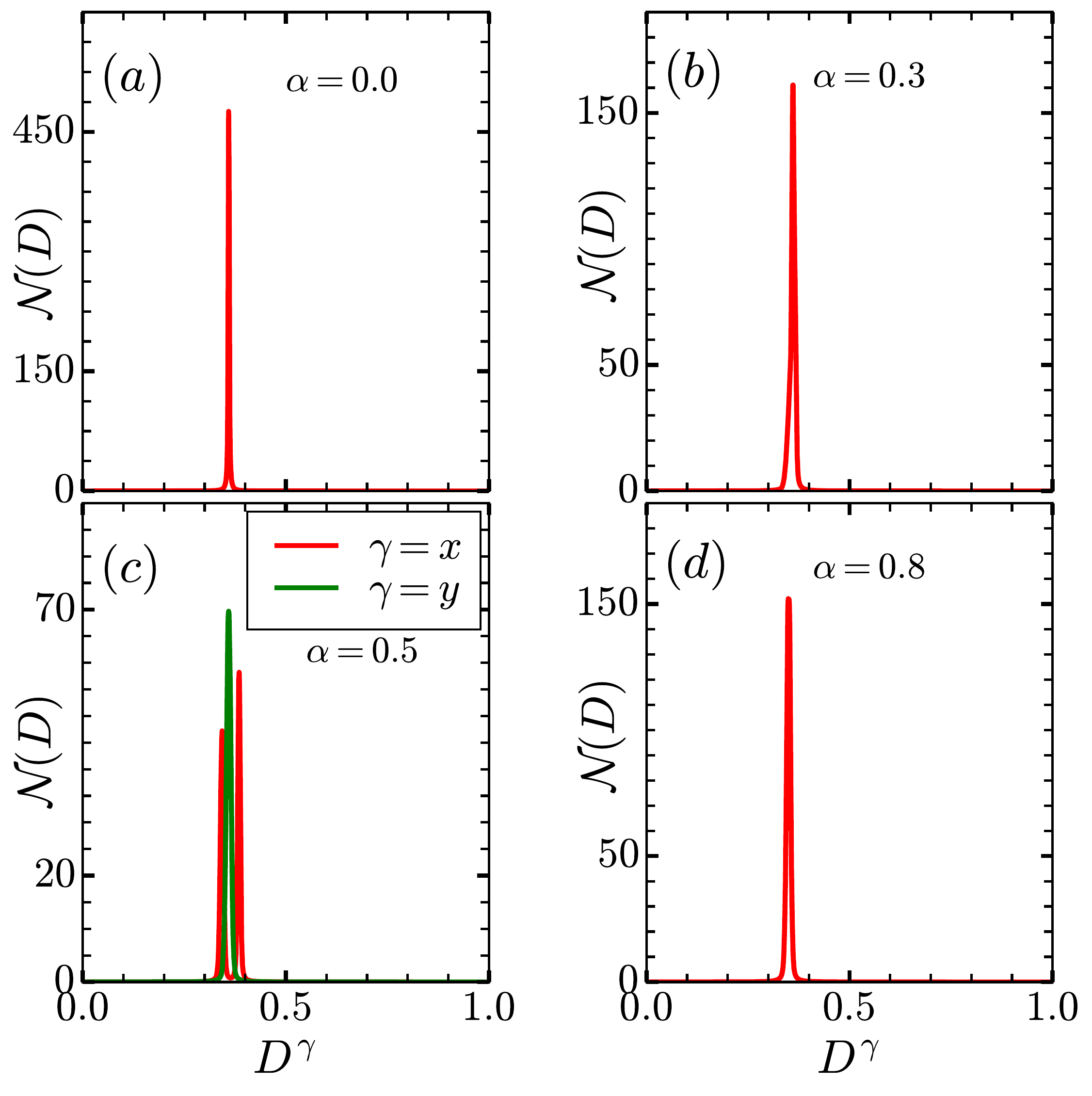}
	\caption{Distributions of $D^{\gamma}_{ij}$ for different ground states, ($a$) Ferromagnet, ($b$) Classical Spin Liquid, ($c$) Single-Q spiral and ($d$) diagonal-Flux, obtained from exact diagonalization of Rashba double exchange Hamiltonian for $N=40^{2}$ at electron filling density of $n=0.3$. }
	\label{fig:DOD}
\end{figure}

\noindent
The density of $D^{\gamma}_{ij}$s is shown in Fig. (\ref{fig:DOD}) for different values of $\alpha$. We find that $D^{\gamma}_{ij}$ is independent of $ij$ for most of the ground states. This justifies the use of a single coupling constant in the effective spin Hamiltonian. For the spiral state with wave vector $(0,q)$ we find a slight separation of scales between $D^{x}_{ij}$ and $ D^{y}_{ij}$. This difference is expected to further re-enforce the stability of the $(0,q)$ spiral states.

\section{Origin of classical spin liquid (CSL) behavior}
In this section we provide a simple description of CSL states observed in the region $0.15 \leq \alpha \leq 0.34$. A careful look at the form of the Hamiltonian Eq. (\ref{eq:ESH}) suggests that for small values of $\alpha$, terms proportional to $\lambda^2$ may be ignored. The only non-trivial effect then comes from terms proportional to $t\lambda$. These terms prefer spiral states with competing orientations of the spiral planes. Along $x$-direction, a spiral in $xz$ plane is preferred and along $y$-direction a spiral in $yz$ plane is preferred.
This motivates us to construct the following variational ansatz where the plane of the spiral is one of the variational parameters:

\begin{eqnarray}
S_i^x &=& S_0 \sin(\boldsymbol{q}.\boldsymbol{r}_i)\cos(\Phi_p), \nonumber \\
S_i^y &=& S_0 \sin(\boldsymbol{q}.\boldsymbol{r}_i)\sin(\Phi_p),\nonumber \\
S_i^z &=& S_0 \cos(\boldsymbol{q}.\boldsymbol{r}_i).
\label{eq:vari_ansatz}
\end{eqnarray}
\noindent
In the above, $S_0$ is the unit magnitude of the classical spin vectors, $\Phi_p$ is the orientation of the spiral plane ($\Phi_p = 0$ for $xz$ plane and $\Phi_p = \frac{\pi}{2}$ for $yz$ plane) and $\boldsymbol{q} = q(\cos\beta,\sin\beta)$ is the spiral wave-vector.
In the CSL state, we find that the energy of a spiral is independent of the spiral plane angle $\Phi_p$, provided the wave-vector angle $\beta$ is related to $\Phi_p$ via $\beta - \Phi_p = \pi$. This explains the stability of filamentary domain wall structure in the CSL regime: the domain walls can freely reorient as long as the spiral plane also undergoes a reorientation in such a way that the spiral plane is oriented perpendicular to the local orientation of the domain wall.

\begin{figure}[t]
	\begin{center}
		\includegraphics[width=0.45 \columnwidth,angle=0,clip=true]{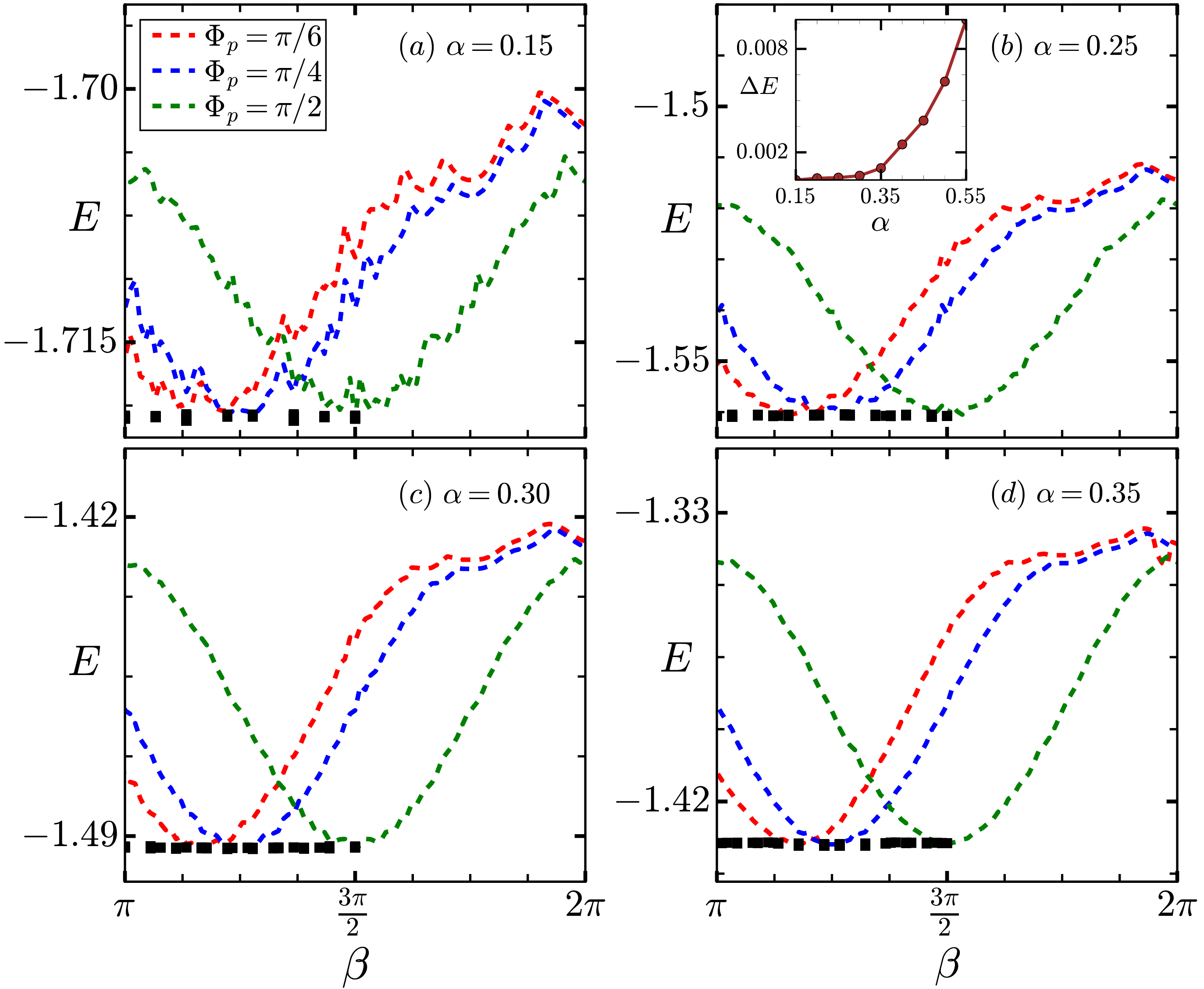}
		\caption{Energy per site $E$ as a function of wave-vector direction $\beta$, for 
			($a$) $\alpha$ = 0.15,
			($b$) $\alpha$ = 0.25,
			($c$) $\alpha$ = 0.30 and
			($d$) $\alpha$ = 0.35,
			obtained for states defined via variational ansatz Eq. (\ref{eq:vari_ansatz}). Energy is minimized over the magnitude $q$ of $\boldsymbol{q}$. Square symbols represent the minimum value, $E_{min}$, of $E$ for each choice of $\Phi_p$. Inset in panel ($b$) shows the variation with $\alpha$ of the width $\Delta E$ of $E_{\text{min}}$. 
		}
		\label{fig:vari_E vs beta}
	\end{center}
	
\end{figure}

In order to quantify this degeneracy of spiral states, we define $\Delta E = \max[E_{min}(\Phi_p)] - \min[E_{min}(\Phi_p)]$. $E_{min}(\Phi_p)$ represents the minimum energy obtained for a given orientation of the spiral plane, marked by a square symbol Fig. \ref{fig:vari_E vs beta}. Exact degeneracy is characterized by $\Delta E = 0$. We show the variation of $\Delta E$ with the coupling constant $\alpha$ as an inset in Fig. \ref{fig:vari_E vs beta} ($b$). The degree of degeneracy clearly reduces near $\alpha=0.35$, which coincides with the crossover point between CSL and SQ spiral states.

\end{document}